\documentclass[11pt]{article}

\usepackage{graphics}
\graphicspath{{./}{figures/}}
\usepackage{cite}
\usepackage{axodraw}
\usepackage{rotating}
\usepackage{url}
\usepackage{xspace}
\usepackage{color}
\usepackage{ulem}
\usepackage{multirow}
\makeatletter
\if@twoside \oddsidemargin -17pt \evensidemargin 00pt
\else \oddsidemargin 00pt \evensidemargin 00pt
\fi
\expandafter\ifx\csname mathrm\endcsname\relax\def\mathrm#1{{\rm #1}}\fi

\renewcommand{\theequation}{\thesection.\arabic{equation}}
\newcounter{saveeqn}

\@addtoreset{equation}{section}

\makeatother

\def\nl{\nonumber\\}

\def\beq{\begin{equation}}
\def\eeq{\end{equation}}
\def\beqar{\begin{eqnarray}}
\def\eeqar{\end{eqnarray}}
\def\barr#1{\begin{array}{#1}}
\def\earr{\end{array}}
\def\bfi{\begin{figure}}
\def\efi{\end{figure}}
\def\btab{\begin{table}}
\def\etab{\end{table}}
\def\bce{\begin{center}}
\def\ece{\end{center}}

\def\text{\textstyle}
\def\arraystretch{1.0}

\def\ga{\gamma}
\def\de{\delta}

\def\si{\sigma}
\def\Ga{\Gamma}

\def\refeq#1{\mbox{(\ref{#1})}}
\def\reffi#1{\mbox{Fig.~\ref{#1}}}

\def\refta#1{\mbox{Table~\ref{#1}}}

\def\refse#1{\mbox{Section~\ref{#1}}}

\def\citere#1{\mbox{Ref.~\cite{#1}}}
\def\citeres#1{\mbox{Refs.~\cite{#1}}}

\newcommand{\GeV}{\unskip\,\mathrm{GeV}}

\newcommand{\TeV}{\unskip\,\mathrm{TeV}}

\def\mathswitch#1{\relax\ifmmode#1\else$#1$\fi}
\def\mathswitchr#1{\relax\ifmmode{\mathrm{#1}}\else$\mathrm{#1}$\fi}
\def\mathswitchit#1{\relax\ifmmode{#1}\else$#1$\fi}

\newcommand{\PW}{\mathswitchr W}
\newcommand{\PZ}{\mathswitchr Z}
\newcommand{\Pg}{\mathswitchr g}
\newcommand{\PH}{\mathswitchr H}
\newcommand{\Pd}{\mathswitchr d}
\newcommand{\Pu}{\mathswitchr u}
\newcommand{\Ps}{\mathswitchr s}
\newcommand{\Pb}{\mathswitchr b}
\newcommand{\Pc}{\mathswitchr c}
\newcommand{\Pt}{\mathswitchr t}
\newcommand{\Pp}{\mathswitchr p}
\newcommand{\Pq}{\mathswitchit q}
\newcommand{\Pep}{\mathswitchr {e^+}}
\newcommand{\Pem}{\mathswitchr {e^-}}
\newcommand{\MW}{\mathswitch {M_{\PW}}}

\newcommand{\MZ}{\mathswitch {M_{\PZ}}}

\newcommand{\GZ}{\mathswitch {\Gamma_{\PZ}}}

\newcommand{\GF}{\mathswitch {G_\mu}}

\newcommand{\alphas}{\alpha_{\mathrm{s}}}
\newcommand{\gs}{g_{\mathrm{s}}}
\newcommand{\pT}{p_{\rm T}}

\def\ie{i.e.\ }

\newcommand{\ps}{p\hspace{-0.42em}/}

\newcommand{\ord}{{\cal O}}

\newcommand{\ri}{{\mathrm{i}}}

\newcommand{\rd}{{\mathrm{d}}}

\newcommand{\M}{{\cal {M}}}

\newcommand{\EW}{{\mathrm{EW}}}

\newcommand{\OS}{{\mathrm{OS}}}

\newcommand{\SM}{{\mathrm{SM}}}

\newcommand{\LO}{{\mathrm{LO}}}
\newcommand{\NLO}{{\mathrm{NLO}}}

\newcommand{\NC}{N_{\mathrm{c}}}

\def\Re{\mathop{\mathrm{Re}}\nolimits}

\def\draftdate{\relax}
\def\mda{\relax}
\def\mua{\relax}
\def\mla{\relax}
\def\draft{
\def\thtystars{******************************}
\def\sixtystars{\thtystars\thtystars}
\typeout{}
\typeout{\sixtystars**}
\typeout{* Draft mode!
         For final version remove \protect\draft\space in source file *}
\typeout{\sixtystars**}
\typeout{}
\def\draftdate{\today}
\def\mua{\marginpar[\boldmath\hfil$\uparrow$]%
                   {\boldmath$\uparrow$\hfil}%
                    \typeout{marginpar: $\uparrow$}\ignorespaces}
\def\mda{\marginpar[\boldmath\hfil$\downarrow$]%
                   {\boldmath$\downarrow$\hfil}%
                    \typeout{marginpar: $\downarrow$}\ignorespaces}
\def\mla{\marginpar[\boldmath\hfil$\rightarrow$]%
                   {\boldmath$\leftarrow $\hfil}%
                    \typeout{marginpar: $\leftrightarrow$}\ignorespaces}
\def\Mua{\marginpar[\boldmath\hfil$\Uparrow$]%
                   {\boldmath$\Uparrow$\hfil}%
                    \typeout{marginpar: $\Uparrow$}\ignorespaces}
\def\Mda{\marginpar[\boldmath\hfil$\Downarrow$]%
                   {\boldmath$\Downarrow$\hfil}%
                    \typeout{marginpar: $\Downarrow$}\ignorespaces}
\def\Mla{\marginpar[\boldmath\hfil$\Rightarrow$]%
                   {\boldmath$\Leftarrow $\hfil}%
                    \typeout{marginpar: $\Leftrightarrow$}\ignorespaces}
\overfullrule 5pt
\oddsidemargin -15mm
\marginparwidth 29mm
}

\makeatletter

\def\eqnarray{\stepcounter{equation}\let\@currentlabel=\theequation
\global\@eqnswtrue
\global\@eqcnt\z@\tabskip\@centering\let\\=\@eqncr
$$\halign to \displaywidth\bgroup\hskip\@centering
  $\displaystyle\tabskip\z@{##}$\@eqnsel&\global\@eqcnt\@ne
  \hskip 2\arraycolsep \hfil${##}$\hfil
  &\global\@eqcnt\tw@ \hskip 2\arraycolsep $\displaystyle\tabskip\z@{##}$\hfil
   \tabskip\@centering&\llap{##}\tabskip\z@\cr}
\def\appendix{\par
 \setcounter{section}{0} \setcounter{subsection}{0}
 \def\thesection{\Alph{section}}}

\def\dsl{\mathpalette\make@slash}
\def\make@slash#1#2{\setbox\z@\hbox{$#1#2$}%
  \hbox to 0pt{\hss$#1/$\hss\kern-\wd0}\box0}

\newcommand{\Pbbar}{{\bar{\Pb}}}
\newcommand{\Pdbar}{{\bar{\Pd}}}
\newcommand{\Pubar}{{\bar{\Pu}}}

\newcommand{\RL}{\mathrm{L}}

\makeatother

\newcommand{\recola}{{\sc Recola}}

\begin{document}

\thispagestyle{empty}
\def\thefootnote{\fnsymbol{footnote}}
\setcounter{footnote}{1}
\null
\draftdate\hfill 
\\
\vskip 1cm
\begin{center}
  {\Large \boldmath{\bf Recursive generation of one-loop amplitudes\\
    in the Standard Model}
\par} \vskip 2.5em
{\large
{\sc S.~Actis$^{1}$, A.~Denner$^{2}$, L.~Hofer$^{2}$, A.~Scharf$^{2}$, 
S.~Uccirati$^{3}$}\\[3ex]
{\normalsize \it
$^1$Paul Scherrer Institut, W\"urenlingen und Villigen,\\
CH-5232 Villigen PSI, Switzerland}\\[1ex]
{\normalsize \it
$^2$Universit\"at W\"urzburg, 
Institut f\"ur Theoretische Physik und Astrophysik, \\
D-97074 W\"urzburg, Germany}\\[1ex]
{\normalsize \it
$^3$Universit\`a di Torino, Dipartimento di Fisica, Italy\\
INFN, Sezione di Torino, Italy}\\[1ex]
}
\par \vskip 1em
\end{center}\par
\vfill \vskip .0cm \vfill {\bf Abstract:} \par 

We introduce the computer code \recola~for the recursive generation of
tree-level and one-loop amplitudes in the Standard Model.  Tree-level
amplitudes are constructed using off-shell currents instead of Feynman
diagrams as basic building blocks. One-loop amplitudes are represented
as linear combinations of tensor integrals whose coefficients are
calculated similarly to the tree-level amplitudes by recursive
construction of loop off-shell currents.  We introduce a novel
algorithm for the treatment of colour, assigning a colour structure to
each off-shell current which enables us to recursively construct the
colour structure of the amplitude efficiently.  \recola~is interfaced
with a tensor-integral library and provides complete one-loop Standard
Model amplitudes including rational terms and counterterms.  As a
first application we consider $\PZ+2\,$jets production at the LHC and
calculate with \recola\ the next-to-leading-order electroweak
corrections to the dominant partonic channels.
\par
\vskip 1cm
\noindent
November 2012
\par
\null
\setcounter{page}{0}
\clearpage
\def\thefootnote{\arabic{footnote}}
\setcounter{footnote}{0}

\section{Introduction}

The study of the mechanism of electroweak (EW) symmetry breaking and
the search for physics beyond the Standard Model (SM) is the primary
goal of the Large Hadron Collider (LHC) and the corresponding
experiments.  With the discovery of a bosonic resonance with a mass of
around $125\GeV$ {important progress has been achieved.} Still it
remains an open question if this resonance is the SM Higgs boson and
if there are phenomena of new physics at the TeV scale. Evidence for a
discovery of new particles and the precise determination of their
masses and couplings on the one hand as well as the establishment of
exclusion limits on the other hand are achieved by sophisticated
experimental analyses, capable of highlighting a small signal on a
huge background. For the interpretation of the data a precise
knowledge of the background is essential. {This} often relies on
theoretical descriptions, sometimes also on data-driven estimations
where the extrapolation to the signal region is based on theoretical
distributions. A sound comparison of experimental signals with
theoretical predictions allowing precise tests of the SM (or of
theories beyond) requires high precision from both experiment and
theory.

Theoretical predictions at leading order (LO) in perturbation theory
are usually insufficient to match the experimental precision.  At a
hadron collider QCD corrections are indispensable, but also EW
corrections can have an important impact.  {For instance,} for
Higgs-boson production in vector-boson fusion, EW and QCD corrections
are of the same order of magnitude~\cite{Ciccolini:2007ec}.
Moreover, the high energies attained by the LHC allow to collect data
in phase-space regions where the effects of logarithms of EW origin
become sizeable.  The high centre-of-mass energies available at the
LHC generate a lot of events with many particles in the final state.
Therefore a proper theoretical description of LHC physics requires
next-to-leading-order (NLO) computations of multiparticle processes
(with five, six, or more external legs) in the full SM (including EW
corrections).

In the past years many groups have concentrated their efforts to make
such calculations feasible.  New techniques have been proposed, mainly
for the computation of one-loop virtual corrections which are
considered the bottleneck of NLO calculations.  In the standard
approach based on Feynman diagrams the major problem is caused by huge
algebraic expressions appearing in the computation of the virtual
amplitudes.  The development of techniques based on Generalised
Unitarity \cite{Bern:1994zx, Bern:1994cg, Britto:2004nc,
  Ossola:2006us,Ossola:2007bb, Ellis:2007br,Giele:2008ve,
  Ellis:2008ir} allowed a change of perspective.  The formal starting
point of these methods is the general decomposition of one-loop
amplitudes as linear combinations of scalar integrals, as obtained
from the standard Passarino--Veltman reduction
\cite{Passarino:1978jh}.  The computation of the coefficients of the
scalar functions is then reduced to the calculation of tree-level
amplitudes by means of cutting equations.  The simplicity of these
relations allowed the automation of NLO QCD computations leading to
the development of computer programs \cite{Berger:2008sj,
  Giele:2008bc, Lazopoulos:2008ex, Giele:2009ui, Badger:2010nx,
  Hirschi:2011pa, Bevilacqua:2011xh, Cullen:2011ac} which enabled the
calculations of many QCD processes~\cite{Berger:2009ep,
  KeithEllis:2009bu,Berger:2009zg, Bevilacqua:2009zn, Melnikov:2009wh,
  Bevilacqua:2010ve, Berger:2010vm, Melia:2010bm, Berger:2010zx,
  Bevilacqua:2010qb, Melia:2011dw, Frederix:2011qg, Ita:2011wn,
  Bevilacqua:2011aa, Bern:2011ep, Greiner:2012im,
  Bevilacqua:2012em,Badger:2012pf} of the Les Houches priority
list~\cite{AlcarazMaestre:2012vp}.

More recently, based on ideas of Soper \cite{Soper:1999xk} purely
numerical methods for the calculation of one-loop QCD amplitudes have
been put forward \cite{Becker:2010ng}.  They are based on subtracting
the soft, collinear, and ultraviolet divergences of one-loop
amplitudes and performing the loop integration of the remaining finite
integrals numerically after {suitably} deforming the integration
contours in the complex space. These methods do not rely on
Feynman graphs and have been proven to work for multi-parton
amplitudes \cite{Becker:2011vg}.

The success of the new methods did, however, not supersede the
traditional diagrammatic approach.  The Generalised Unitarity methods
still suffer from the numerical instabilities characteristic of the
Passarino--Veltman reduction. These are overcome by computing the
amplitude in quadruple or multiple precision at critical phase-space
points, reducing however the CPU
efficiency~\cite{Ossola:2007ax,Mastrolia:2010nb}.  While rescue
solutions in this context have been proposed in
\citeres{Heinrich:2010ax,Pittau:2010tk}, it is well-known that
numerical instabilities can be avoided by constructing the amplitude
as a linear combination of tensor integrals, which is the natural
representation in the Feynman-diagrammatic approach.  Various groups
\cite{Cullen:2011ac,Ferroglia:2002mz, Denner:2002ii, Binoth:2005ff,
  Denner:2005nn} have developed efficient computational techniques for
the calculation of the tensor integrals that avoid numerical
instabilities.  Using these improved methods in the
Feynman-diagrammatic approach yields more than competitive numerical
codes. In particular the methods of~\citeres{Denner:2002ii,
  Denner:2005nn} have been successfully applied to the calculation of
both EW corrections~\cite{Bredenstein:2006ha, Ciccolini:2007ec,
  Denner:2009gj, Denner:2011vu, Denner:2011id} and QCD
corrections~\cite{Bredenstein:2009aj, Denner:2010jp,
  Campanario:2011ud,Reina:2011mb}.

Recently the diagrammatic approach has been further boosted by the
advent of the {\sc OpenLoops} algorithm~\cite{Cascioli:2011va}.
Organising the diagrams into cut-opened topologies, the coefficients
of the tensor integrals are recursively built with tree-level-like
techniques; the efficiency is increased by pinch identities that
relate higher-point loop diagrams to pre-computed lower-point
diagrams. Since the method is based on individual topologies, colour
factorises and is treated algebraically. The present implementation of
{\sc OpenLoops} can handle NLO QCD corrections to any Standard Model
process.

An interesting hybrid method, proposed by Andreas van Hameren in
\citere{vanHameren:2009vq} for evaluating one-loop gluonic
amplitudes, combines stable and universal tensor-reduction methods for
loop integrals with a basic result of Generalised Unitarity, namely
the reduction of the computation of a one-loop amplitude to that of a
set of tree-level amplitudes.  The technique relies on the
representation of the amplitude in terms of tensor integrals, whose
coefficients are computed recursively~\cite{Berends:1987me} without
resorting to Feynman diagrams at any stage.

In this paper we present \recola, a generator of SM one-loop (and
tree) amplitudes.  It is based on an algorithm which implements
recursion relations for the computation of the coefficients of the
tensor integrals.  The goal is to combine the {efficiency} of the
numerically stable tensor-integral reduction with the automation made
possible by a completely recursive non-diagrammatic approach.

As a first application of \recola~we have calculated EW corrections
to $\Pp\Pp \to \PZ+2\,$jets.  Due to its large cross section and
similar signatures this process provides a major background for
Higgs-boson production in vector-boson fusion kinematics.  The
dominant NLO QCD corrections of $\ord(\alphas^3\alpha)$ have been
investigated in \citeres{Campbell:2002tg,Campbell:2003hd}, while a
subset of EW $\PZ+2\,$jets production has been studied at NLO QCD in
\citere{Oleari:2003tc}.  {Here, we calculate the EW} corrections of
$\ord(\alphas^2\alpha^2)$ to the dominant partonic channels for
$\PZ+2\,$jets production.  

This paper is organised as follows: in \refse{bulba} we present the
algorithm {for the construction of amplitudes,} introducing first our
implementation of recursion relations at tree level (\refse{tree}) and
discussing in \refse{loop} its generalisation to one-loop amplitudes.
After some remarks on the computation of the rational terms and
counterterms (\refse{R2&CT}), we describe in \refse{colour} a new
algorithm for the treatment of colour.  The setup of our calculation
for $\PZ+2\,$jets production is detailed in \refse{ppZjj details}, and
numerical results are discussed in \refse{ppZjj results}.  Finally,
\refse{conclusions} contains our conclusions.

\section{\recola: REcursive Computation of One-Loop Amplitudes}
\label{bulba}
\recola\ is a code written in {\sc Fortran90} for the computation of
tree and one-loop scattering amplitudes in the SM, based on recursion
relations.
\subsection{Tree-level recursion relations}
\label{tree}
The tree-level recursive algorithm is inspired by the Dyson--Schwinger
equations~{\cite{Dyson:1949ha,Schwinger:1951ex,Schwinger:1951hq}} and
follows closely the strategy of {\sc
  Helac}~\cite{Kanaki:2000ey,Papadopoulos:2005ky,Cafarella:2007pc},
using off-shell currents as basic building blocks.

Let us consider a process with $E$ external legs, and
select a particle $P$ of the model%
\footnote{Here all particles of the model have to be taken into
  account, also unphysical ones like would-be Goldstone bosons.}
together with a sub-set of $n$ external legs. The off-shell current
$w(P,\{n\})$ is {then} defined as the sum of all Feynman sub-graphs
which generate $P$ combining the selected $n$ external particles:
\beq
w(P,\{n\})
\quad=\qquad\vcenter{\hbox{
\begin{picture}(55,30)(-5,-15)
\Line(20,0)(40,0)
\GCirc(40,0){1.5}{0}
\Line(0,15)(20,0)
\Line(0,-15)(20,0)
\GCirc(20,0){10}{.8}
\DashCArc(20,0)(22,155,205){2}
\Text(-9,-3)[cb]{$n$} 
\Text(42,6)[cb]{$P$} 
\end{picture}}}.
\label{current}
\eeq
Here the shaded bubble pictorially represents all possible
sub-graphs, and the dot indicates the off-shell part of the current.
If according to the Feynman rules of the theory {a sub-set} of $n$
external particles cannot generate $P$, the corresponding current
vanishes.  For $n>1$ the propagator of $P$ is included in the
definition and the current is called ``internal".  A current
generated by only one ($n=1$) external particle $P'$ is called
``external"; if $P=P'$, it coincides with the wave function of the
particle, otherwise it vanishes.

In a theory with tri- and quadri-linear couplings only, the internal
off-shell currents can be constructed using recursively the
Dyson--Schwinger equations:
\beq
\begin{picture}(55,40)(-5,-2)
\Line(20,0)(40,0)
\GCirc(40,0){1.5}{0}
\Line(0,15)(20,0)
\Line(0,-15)(20,0)
\GCirc(20,0){10}{.8}
\DashCArc(20,0)(22,155,205){2}
\Text(-9,-3)[cb]{$n$} 
\Text(42,5)[cb]{$P$} 
\end{picture}
\quad
=
\quad
\sum_{\{i\},\{j\}}^{i+j=n}\;
\sum_{P_i,P_j}\;
\begin{picture}(65,40)(-10,-2)
\Line(0,25)(15,15)
\Line(0,5)(15,15)
\Line(15,15)(35,0)
\GCirc(15,15){7.5}{.8}
\DashCArc(15,15)(16,160,200){2}
\Text(31,8)[cb]{\scriptsize $P_{\!i}$}
\Text(-6,12)[cb]{$i$} 
\Line(0,-5)(15,-15)
\Line(0,-25)(15,-15)
\Line(15,-15)(35,0)
\GCirc(15,-15){7.5}{.8}
\DashCArc(15,-15)(16,160,200){2}
\Text(31,-17)[cb]{\scriptsize $P_{\!j}$}
\Text(-6,-18)[cb]{$j$} 
\SetWidth{.9}
\Line(35,0)(50,0)
\SetWidth{.5}
\BBoxc(34,0)(5,5)
\GCirc(50,0){1.5}{0}
\Text(52,5)[cb]{$P$} 
\end{picture}
\quad
+
\;
\sum_{\{i\},\{j\},\{k\}}^{i+j+k=n}\;
\sum_{P_{\!i},P_{\!j},P_{\!k}}\;
\begin{picture}(70,40)(-10,-2)
\Line(0,35)(15,25)
\Line(0,15)(15,25)
\Line(15,25)(45,0)
\GCirc(15,25){7.5}{.8}
\DashCArc(15,25)(16,160,200){2}
\Text(-6,22)[cb]{$i$} 
\Text(31,18)[cb]{\scriptsize $P_{\!i}$}
\Line(0,10)(15,0)
\Line(0,-10)(15,0)
\Line(15,0)(45,0)
\GCirc(15,0){7.5}{.8}
\DashCArc(15,0)(16,160,200){2}
\Text(-6,-3)[cb]{$j$} 
\Text(27,1)[cb]{\scriptsize $P_{\!j}$}
\Line(0,-15)(15,-25)
\Line(0,-35)(15,-25)
\Line(15,-25)(45,0)
\GCirc(15,-25){7.5}{.8}
\DashCArc(15,-25)(16,160,200){2}
\Text(-6,-28)[cb]{$k$} 
\Text(31,-27)[cb]{\scriptsize $P_{\!k}$}
\SetWidth{.9}
\Line(45,0)(60,0)
\SetWidth{.5}
\BBoxc(43,0)(5,5)
\GCirc(60,0){1.5}{0}.
\Text(60,5)[cb]{$P$}
\end{picture}
\;\;\;.
\label{recursive}
\eeq

\vspace{.7cm}
\noindent
Each term of the sums represents a ``branch", which
is obtained by multiplication of the generating currents
$w(P_{\!i},\{i\})$, $w(P_{\!j},\{j\})$ [and $w(P_{\!k},\{k\})$ for
the second term of \refeq{recursive}] with the interaction vertex,
marked by the small box, and the propagator of $P$, marked by the
thick line.  Branches have to be built for all possible generating
currents formed by sub-sets $\{i\}$, $\{j\}$ (and $\{k\}$) of the $n$
external particles such that $i+j\;(+k)=n$.  The recursive
construction of the amplitude can be efficiently implemented using
lower-multiplicity off-shell currents as seeds for the numerical
evaluation of higher-multiplicity ones.  First, all possible currents
with two external legs ($n=2$) are calculated combining through a
tri-linear coupling pairs of external currents only.  Next the
currents with three external particles ($n=3$) are generated summing
the branches with three external currents linked through a
quadri-linear coupling and those with one external current and one of
the calculated internal currents with two external legs, linked
through a tri-linear coupling.  Analogously the currents with $n=4$
are then computed combining two or three currents with $n=1$, $n=2$
and $n=3$, and so on.  As a consequence of the summation in
\refeq{recursive}, the current {$w(P,\{n\})$} depends on the particle
{$P$} and on the {set of generating external particles $\{n\}$} but
not on the particular way these particles have been combined in order
to get $w(P,\{n\})$.  In this respect, working with off-shell currents
rather than Feynman diagrams allows to avoid recomputing
{identical} sub-graphs contributing to {different} diagrams, since
each current is computed just once; furthermore, the summation {in}
\refeq{recursive} reduces the number of generated objects which are
passed to the next step of the recursion.

In order to calculate the amplitude for a tree-level process $A\to B$
we first use crossing symmetry {to switch to incoming particles} and
consider the corresponding process $A+\overline{B}\to0$, where
$\overline{B}$ is charge conjugate to $B$.  Next we choose one of the
external legs, for definiteness the $E^{\rm th}$ leg, to close the
construction of the amplitude. Then all currents resulting from the
other $E-1$ external legs are recursively constructed.  In the last
step of the procedure, \ie for $n=E-1$, we require the generated
particle $P$ to be the selected $E^{\rm th}$ external particle.
{As} a consequence this last current is unique, and the amplitude
$\M$ is obtained multiplying with the inverse of its propagator and
with the wave function of the selected $E^{\rm th}$ particle (which
coincides with the $E^{\rm th}$ external current):

\beq
\M \;=\vcenter{\hbox{
\begin{picture}(80,30)(-35,-15)
\Line(20,0)(40,0)
\GCirc(40,0){1.5}{0}
\Line(0,15)(20,0)
\Line(0,-15)(20,0)
\GCirc(20,0){10}{.8}
\DashCArc(20,0)(22,155,205){2}
\Text(-17,-3)[cb]{$E\!-\!1$} 
\end{picture}}}
\times
(\mbox{propagator})^{-1}
\times\vcenter{\hbox{
\begin{picture}(40,10)(-5,-5)
\Line(0,0)(20,0)
\GCirc(0,0){1.5}{0}
\Text(30,-3)[cb]{$E^{\rm}$} 
\end{picture}\,\,\,.}}
\label{last step}
\eeq

The recursive evaluation of the currents begins with the external 
currents ($n=1$), which, for colourless particles of a given polarisation 
$\lambda$ and momentum $p$, are given by the corresponding wave functions:
\vspace{-1ex}
\begin{equation}\vcenter{\hbox{
\begin{picture}(31,30)(-5,-15)
\ArrowLine(0,0)(20,0)
\GCirc(20,0){1.5}{0}
\Text(-5,-3)[cb]{$\lambda$}
\LongArrow(5,-7)(15,-7)\Text(10,-17)[cb]{$p$}
\end{picture}}}
= \,
u_\lambda(p),
\qquad\vcenter{\hbox{
\begin{picture}(31,10)(-5,-5)
\ArrowLine(20,0)(0,0)
\GCirc(20,0){1.5}{0}
\Text(-5,-3)[cb]{$\lambda$}
\LongArrow(5,-7)(15,-7)\Text(10,-17)[cb]{$p$}
\end{picture}}}
= \,
\bar{v}_\lambda(p),
\qquad\vcenter{\hbox{
\begin{picture}(31,10)(-5,-5)
\Photon(0,0)(20,0){2.5}{3}
\GCirc(20,0){1.5}{0}
\Text(-5,-3)[cb]{$\lambda$}
\LongArrow(5,-7)(15,-7)\Text(10,-17)[cb]{$p$}
\end{picture}}}
= \;
\epsilon_\lambda(p),
\qquad\vcenter{\hbox{
\begin{picture}(31,10)(-5,-5)
\DashLine(0,0)(20,0){4}
\GCirc(20,0){1.5}{0}
\LongArrow(5,-7)(15,-7)\Text(10,-17)[cb]{$p$}
\end{picture}}}
= \;
1\,.
\end{equation}
The explicit expressions for the spinors $u_\lambda(p)$,
$\bar{u}_\lambda(p)$, $v_\lambda(p)$, $\bar{v}_\lambda(p)$ of fermions
and for the polarisation vectors $\epsilon_\lambda(p)$,
$\epsilon^*_\lambda(p)$ of vector bosons have been coded using
\citere{Hagiwara:1988pp}. For coloured external particles also the
information on colour has to be kept, as we explain in \refse{colour}.

Once the external currents have been numerically evaluated, the
internal ones are built using the Feynman rules of the theory.  For
example, given a pair of external $\Pem$ and $\Pep$ with momenta $p_1$
and $p_2$ and currents $u_{\lambda_1}(p_1)$ and
$\bar{v}_{\lambda_2}(p_2)$, one can generate the internal current of a
photon contracting the two external currents with the QED vertex
$-\ri\,e\,\gamma^\mu$ and the photon propagator
$-\ri\,g_{\mu\nu}/(p_1+p_2)^2$.  Once the particle $P$ of the
off-shell propagator is fixed, in general, several branches contribute
to the same internal current since different combinations of particles
with appropriate interaction vertices can generate the same particle
$P$.  In these cases the contributions can be simply summed up,
according to \refeq{recursive}.

The recursive algorithm is implemented in the code \recola\ through
two steps.

In the first part, the initialisation phase, the currents are
identified by integer numbers which contain all the relevant
non-dynamical information: the particle content ($\Pem$, $\Pep$,
$\gamma$, etc.), the colour information (see \refse{colour}) and an
integer tag number.  The tag number is assigned according to a binary
notation~\cite{Kanaki:2000ey,Caravaglios:1995cd}.  In practice, the
external currents get a label $2^{i-1}$, where $i$ is the $i^{\rm th}$
external leg, \ie $1 \to 1,\;2 \to 2,\; 3 \to 4, \;\dots,\;E\to
2^{E-1}$.  The tag number of an internal current is obtained by a
summation of the integer tags of the external currents
contributing to it.  The binary notation ensures that, given the tag
number of any internal current, the contributing external legs can be
uniquely identified.  Moreover, it reflects the basic property that a
current depends on the generating external particles $\{n\}$, but not
on the particular way these particles have been combined in order to
obtain the current.

The initialisation part of the code builds a skeleton of the
amplitude: all needed off-shell currents are enumerated and, for each
branch, all generating off-shell currents and the generated one are
identified.  This part is run once for all, before giving explicit
values to the momenta of the external particles, i.e.\ before
performing the phase-space integration in a Monte Carlo program.

The second part of the code, the dynamical production phase, uses the
results of the first part to actually compute the amplitude for each
point of the phase-space.  First, the external currents are
numerically computed; then, branch after branch, all internal currents
are recursively evaluated according to the skeleton generated in the
initialisation phase.  Here the code has to be as efficient as
possible in terms of CPU time because it must run on a large grid of
points in phase-space; the computation of all non-dynamical quantities
in the initialisation phase allows to avoid a repetition of those
operations which can be performed once independently of the particular
values of the external momenta.

\subsection{One-loop recursion relations}
\label{loop}
Let us now move to the one-loop case.  After summing all contributing
Feynman graphs ${\cal G}_{{ i}}$ every one-loop amplitude can be
written as a linear combination of tensor integrals:
\beq
\delta\M = \sum_{{ i}}\,{\cal G}_{{ i}} = 
\sum_{j}\,\sum_{r_j}\,c_{\mu_1\cdots\mu_{r_j}}^{(j,r_j,{N_j})}\,
T_{(j,r_j,{N_j})}^{\mu_1\cdots\mu_{r_j}}\,.
\label{tensor splitting}
\eeq
Here the tensor coefficients
$c_{\mu_1\cdots\mu_{r_j}}^{(j,r_j,{N_j})}$ do not depend on the loop
momentum $q$, which is present only in the tensor integrals
\beq
T_{(j,r_j,{N_j})}^{\mu_1\cdots\mu_{r_j}} =
{\frac{(2\pi\mu)^{4-D}}{\ri\pi^2}}
\int
\rd^Dq\, \frac{q^{\mu_1} \cdots q^{\mu_{r_j}}}{D_{j,0}\cdots D_{j,{N_j-1}}},
\qquad
{D_{j,a} = (q+p_{j,a})^2 - m_{j,a}^2.}
\eeq
The index $j$ classifies the different tensor integrals needed for the
process, the integer {$N_j$} {equals} the number of loop
propagators, and $r_j$ ($\,\le {N_j}$ in the 't Hooft--Feynman
gauge) is the rank of the tensor integral.

Leaving aside the computation of the tensor integrals, to be performed
with the preferred technique, we focus here on the tensor coefficients
$c_{\mu_1\cdots\mu_{r_j}}^{(j,r_j,{N_j})}$, which for multi-leg
processes, due to the complexity of the SM, result in long algebraic
expressions in the standard Feynman-diagram approach.  An interesting
idea has been proposed in \citere{vanHameren:2009vq}, where recursive
relations for the tensor coefficients of gluon amplitudes have been
derived for colour-ordered amplitudes of purely gluonic processes. We
have further developed this approach to deal with the full SM.  There
is a clear topological correspondence between a one-loop diagram with
$E$ external legs and a tree diagram with $E+2$ external legs,
obtained after cutting one of the loop lines.  After uniquely fixing
this correspondence, one can compute the tensor coefficients
{$c_{\mu_1\cdots\mu_{r_j}}^{(j,r_j,N_j)}$} with recursion relations
similar to those used for tree amplitudes.

Given a process $A\to B$ at one loop, we consider first the set of all
tree processes $A+\overline{B}+P+\bar{P}\to 0$ for each particle $P$
of the SM.%
\footnote{Here all unphysical particles, in particular also
    Faddeev--Popov ghosts, must be included.} %
However, the set $\{A+\overline{B}+P+\bar{P}\to 0,\,\forall P\in\SM\}$
contains more diagrams than the original one-loop process $A\to B$.
This is due to the fact that we can cut the loop diagram at any of its
loop lines and {that} we can run along the loop clockwise and
counterclockwise.  {Therefore,} we have to fix some rules to
discard the redundant diagrams.  To explain these rules, we can work
without loss of generality in a theory with a single scalar particle
$\phi$ with a tri-linear coupling ({the generalisation to} the
presence of a quadri-linear coupling {is straightforward}).
In such a theory the set $\{A+\overline{B}+P+\bar{P}\to 0,\,\forall
P\in\SM\}$ reduces to one process, i.e.\ $A+\overline{B}
+\phi+\bar{\phi}\to0$.  We use tag numbers for external and internal
currents as explained in \refse{tree} and assign tag numbers $2^E$ and
$2^{E+1}$ to the currents corresponding to the two additional external
legs of $P$ and $\bar{P}$.  These currents are called ``external loop
currents", while the legs of $P$ and $\bar{P}$ are called ``external
loop legs".

Let us first consider the sets of diagrams with three and four
external legs where only external particles enter the loop.  Marking
with a cross the two external loop legs of the trees, we get:
\beq
\scalebox{0.8}{
\begin{picture}(45,30)(-2,-2)
\Line(0,0)(10,0)
\Line(40,-15)(28,-6)
\Line(40,15)(28,6)
\CArc(20,0)(10,0,360)
\Text(-4,-3)[cb]{\small $1$}
\Text(40,-24)[cb]{\small $2$}
\Text(40,18)[cb]{\small $4$}
\end{picture}
}
\to\;\;
\scalebox{0.8}{
\begin{picture}(45,30)(-2,-2)
\Line(0,0)(10,0)
\Line(40,-15)(28,-6)
\Line(40,15)(28,6)
\CArc(20,0)(10,143,37)
\Line(12,15)(12,6)
\Text(12,12.5)[cb]{\scriptsize $\times$}
\Line(28,15)(28,6)
\Text(28,12.5)[cb]{\scriptsize $\times$}
\Text(-4,-3)[cb]{\small $1$}
\Text(45,-18)[cb]{\small $2$}
\Text(45,12)[cb]{\small $4$}
\Text(12,19)[cb]{\small $8$}
\Text(28,19)[cb]{\small $16$}
\end{picture}
}
+
\scalebox{0.8}{
\begin{picture}(45,30)(-2,-2)
\Line(0,0)(10,0)
\Line(40,-15)(28,-6)
\Line(40,15)(28,6)
\CArc(20,0)(10,143,37)
\Line(12,15)(12,6)
\Text(12,12.5)[cb]{\scriptsize $\times$}
\Line(28,15)(28,6)
\Text(28,12.5)[cb]{\scriptsize $\times$}
\Text(-4,-3)[cb]{\small $1$}
\Text(45,-18)[cb]{\small $4$}
\Text(45,12)[cb]{\small $2$}
\Text(12,19)[cb]{\small $8$}
\Text(28,19)[cb]{\small $16$}
\end{picture}
}
+
\scalebox{0.8}{
\begin{picture}(45,30)(-2,-2)
\Line(0,0)(10,0)
\Line(40,-15)(28,-6)
\Line(40,15)(28,6)
\CArc(20,0)(10,143,37)
\Line(12,15)(12,6)
\Text(12,12.5)[cb]{\scriptsize $\times$}
\Line(28,15)(28,6)
\Text(28,12.5)[cb]{\scriptsize $\times$}
\Text(-4,-3)[cb]{\small $2$}
\Text(45,-18)[cb]{\small $1$}
\Text(45,12)[cb]{\small $4$}
\Text(12,19)[cb]{\small $8$}
\Text(28,19)[cb]{\small $16$}
\end{picture}
}
+
\scalebox{0.8}{
\begin{picture}(45,30)(-2,-2)
\Line(0,0)(10,0)
\Line(40,-15)(28,-6)
\Line(40,15)(28,6)
\CArc(20,0)(10,143,37)
\Line(12,15)(12,6)
\Text(12,12.5)[cb]{\scriptsize $\times$}
\Line(28,15)(28,6)
\Text(28,12.5)[cb]{\scriptsize $\times$}
\Text(-4,-3)[cb]{\small $2$}
\Text(45,-18)[cb]{\small $4$}
\Text(45,12)[cb]{\small $1$}
\Text(12,19)[cb]{\small $8$}
\Text(28,19)[cb]{\small $16$}
\end{picture}
}
+
\scalebox{0.8}{
\begin{picture}(45,30)(-2,-2)
\Line(0,0)(10,0)
\Line(40,-15)(28,-6)
\Line(40,15)(28,6)
\CArc(20,0)(10,143,37)
\Line(12,15)(12,6)
\Text(12,12.5)[cb]{\scriptsize $\times$}
\Line(28,15)(28,6)
\Text(28,12.5)[cb]{\scriptsize $\times$}
\Text(-4,-3)[cb]{\small $4$}
\Text(45,-18)[cb]{\small $1$}
\Text(45,12)[cb]{\small $2$}
\Text(12,19)[cb]{\small $8$}
\Text(28,19)[cb]{\small $16$}
\end{picture}
}
+
\scalebox{0.8}{
\begin{picture}(45,30)(-2,-2)
\Line(0,0)(10,0)
\Line(40,-15)(28,-6)
\Line(40,15)(28,6)
\CArc(20,0)(10,143,37)
\Line(12,15)(12,6)
\Text(12,12.5)[cb]{\scriptsize $\times$}
\Line(28,15)(28,6)
\Text(28,12.5)[cb]{\scriptsize $\times$}
\Text(-4,-3)[cb]{\small $4$}
\Text(45,-18)[cb]{\small $2$}
\Text(45,12)[cb]{\small $1$}
\Text(12,19)[cb]{\small $8$}
\Text(28,19)[cb]{\small $16$}
\end{picture}
}\;\;,
\nonumber
\eeq
$$
\begin{picture}(40,10)(-2,-2)
\end{picture}
$$
\begin{eqnarray}
\scalebox{0.75}{
\begin{picture}(40,30)(-2,-2)
\Line(0,-15)(12,-6)
\Line(0,15)(12,6)
\Line(40,-15)(28,-6)
\Line(40,15)(28,6)
\CArc(20,0)(10,0,360)
\Text(0, 18)[cb]{\small $1$}
\Text(0,-24)[cb]{\small $2$}
\Text(40,-24)[cb]{\small $4$}
\Text(40, 18)[cb]{\small $8$}
\end{picture}
}
\,+\!
\scalebox{0.75}{
\begin{picture}(40,30)(-2,-2)
\Line(0,-15)(12,-6)
\Line(0,15)(12,6)
\Line(40,-15)(28,-6)
\Line(40,15)(28,6)
\CArc(20,0)(10,0,360)
\Text(0, 18)[cb]{\small $1$}
\Text(0,-24)[cb]{\small $2$}
\Text(40,-24)[cb]{\small $8$}
\Text(40, 18)[cb]{\small $4$}
\end{picture}
}
\,+\!
\scalebox{0.75}{
\begin{picture}(40,30)(-2,-2)
\Line(0,-15)(12,-6)
\Line(0,15)(12,6)
\Line(40,-15)(28,-6)
\Line(40,15)(28,6)
\CArc(20,0)(10,0,360)
\Text(0, 18)[cb]{\small $1$}
\Text(0,-24)[cb]{\small $8$}
\Text(40,-24)[cb]{\small $2$}
\Text(40, 18)[cb]{\small $4$}
\end{picture}
}
&\;\to\;&
\phantom{{}+}
\scalebox{0.75}{
\begin{picture}(45,30)(-2,-2)
\Line(0,-15)(12,-6)
\Line(0,15)(12,6)
\Line(40,-15)(28,-6)
\Line(40,15)(28,6)
\CArc(20,0)(10,143,37)
\Line(12,15)(12,6)
\Text(12,12.5)[cb]{\scriptsize $\times$}
\Line(28,15)(28,6)
\Text(28,12.5)[cb]{\scriptsize $\times$}
\Text(-4, 12)[cb]{\small $1$}
\Text(-4,-18)[cb]{\small $2$}
\Text(45,-18)[cb]{\small $4$}
\Text(45, 12)[cb]{\small $8$}
\Text(12,19)[cb]{\small $16$}
\Text(28,19)[cb]{\small $32$}
\end{picture}
}
+
\scalebox{0.75}{
\begin{picture}(45,30)(-2,-2)
\Line(0,-15)(12,-6)
\Line(0,15)(12,6)
\Line(40,-15)(28,-6)
\Line(40,15)(28,6)
\CArc(20,0)(10,143,37)
\Line(12,15)(12,6)
\Text(12,12.5)[cb]{\scriptsize $\times$}
\Line(28,15)(28,6)
\Text(28,12.5)[cb]{\scriptsize $\times$}
\Text(-4, 12)[cb]{\small $2$}
\Text(-4,-18)[cb]{\small $4$}
\Text(45,-18)[cb]{\small $8$}
\Text(45, 12)[cb]{\small $1$}
\Text(12,19)[cb]{\small $16$}
\Text(28,19)[cb]{\small $32$}
\end{picture}
}
+
\scalebox{0.75}{
\begin{picture}(45,30)(-2,-2)
\Line(0,-15)(12,-6)
\Line(0,15)(12,6)
\Line(40,-15)(28,-6)
\Line(40,15)(28,6)
\CArc(20,0)(10,143,37)
\Line(12,15)(12,6)
\Text(12,12.5)[cb]{\scriptsize $\times$}
\Line(28,15)(28,6)
\Text(28,12.5)[cb]{\scriptsize $\times$}
\Text(-4, 12)[cb]{\small $4$}
\Text(-4,-18)[cb]{\small $8$}
\Text(45,-18)[cb]{\small $1$}
\Text(45, 12)[cb]{\small $2$}
\Text(12,19)[cb]{\small $16$}
\Text(28,19)[cb]{\small $32$}
\end{picture}
}
+
\scalebox{0.75}{
\begin{picture}(45,30)(-2,-2)
\Line(0,-15)(12,-6)
\Line(0,15)(12,6)
\Line(40,-15)(28,-6)
\Line(40,15)(28,6)
\CArc(20,0)(10,143,37)
\Line(12,15)(12,6)
\Text(12,12.5)[cb]{\scriptsize $\times$}
\Line(28,15)(28,6)
\Text(28,12.5)[cb]{\scriptsize $\times$}
\Text(-4, 12)[cb]{\small $8$}
\Text(-4,-18)[cb]{\small $1$}
\Text(45,-18)[cb]{\small $2$}
\Text(45, 12)[cb]{\small $4$}
\Text(12,19)[cb]{\small $16$}
\Text(28,19)[cb]{\small $32$}
\end{picture}
}
\nonumber\\[.4cm]{}
&&
{}+
\scalebox{0.75}{
\begin{picture}(45,30)(-2,-2)
\Line(0,-15)(12,-6)
\Line(0,15)(12,6)
\Line(40,-15)(28,-6)
\Line(40,15)(28,6)
\CArc(20,0)(10,143,37)
\Line(12,15)(12,6)
\Text(12,12.5)[cb]{\scriptsize $\times$}
\Line(28,15)(28,6)
\Text(28,12.5)[cb]{\scriptsize $\times$}
\Text(-4, 12)[cb]{\small $8$}
\Text(-4,-18)[cb]{\small $4$}
\Text(45,-18)[cb]{\small $2$}
\Text(45, 12)[cb]{\small $1$}
\Text(12,19)[cb]{\small $16$}
\Text(28,19)[cb]{\small $32$}
\end{picture}
}
+
\scalebox{0.75}{
\begin{picture}(45,30)(-2,-2)
\Line(0,-15)(12,-6)
\Line(0,15)(12,6)
\Line(40,-15)(28,-6)
\Line(40,15)(28,6)
\CArc(20,0)(10,143,37)
\Line(12,15)(12,6)
\Text(12,12.5)[cb]{\scriptsize $\times$}
\Line(28,15)(28,6)
\Text(28,12.5)[cb]{\scriptsize $\times$}
\Text(-4, 12)[cb]{\small $1$}
\Text(-4,-18)[cb]{\small $8$}
\Text(45,-18)[cb]{\small $4$}
\Text(45, 12)[cb]{\small $2$}
\Text(12,19)[cb]{\small $16$}
\Text(28,19)[cb]{\small $32$}
\end{picture}
}
+
\scalebox{0.75}{
\begin{picture}(45,30)(-2,-2)
\Line(0,-15)(12,-6)
\Line(0,15)(12,6)
\Line(40,-15)(28,-6)
\Line(40,15)(28,6)
\CArc(20,0)(10,143,37)
\Line(12,15)(12,6)
\Text(12,12.5)[cb]{\scriptsize $\times$}
\Line(28,15)(28,6)
\Text(28,12.5)[cb]{\scriptsize $\times$}
\Text(-4, 12)[cb]{\small $2$}
\Text(-4,-18)[cb]{\small $1$}
\Text(45,-18)[cb]{\small $8$}
\Text(45, 12)[cb]{\small $4$}
\Text(12,19)[cb]{\small $16$}
\Text(28,19)[cb]{\small $32$}
\end{picture}
}
+
\scalebox{0.75}{
\begin{picture}(45,30)(-2,-2)
\Line(0,-15)(12,-6)
\Line(0,15)(12,6)
\Line(40,-15)(28,-6)
\Line(40,15)(28,6)
\CArc(20,0)(10,143,37)
\Line(12,15)(12,6)
\Text(12,12.5)[cb]{\scriptsize $\times$}
\Line(28,15)(28,6)
\Text(28,12.5)[cb]{\scriptsize $\times$}
\Text(-4, 12)[cb]{\small $4$}
\Text(-4,-18)[cb]{\small $2$}
\Text(45,-18)[cb]{\small $1$}
\Text(45, 12)[cb]{\small $8$}
\Text(12,19)[cb]{\small $16$}
\Text(28,19)[cb]{\small $32$}
\end{picture}
}
\nonumber\\[.4cm]{}
&&
{}+
\scalebox{0.75}{
\begin{picture}(45,30)(-2,-2)
\Line(0,-15)(12,-6)
\Line(0,15)(12,6)
\Line(40,-15)(28,-6)
\Line(40,15)(28,6)
\CArc(20,0)(10,143,37)
\Line(12,15)(12,6)
\Text(12,12.5)[cb]{\scriptsize $\times$}
\Line(28,15)(28,6)
\Text(28,12.5)[cb]{\scriptsize $\times$}
\Text(-4, 12)[cb]{\small $1$}
\Text(-4,-18)[cb]{\small $2$}
\Text(45,-18)[cb]{\small $8$}
\Text(45, 12)[cb]{\small $4$}
\Text(12,19)[cb]{\small $16$}
\Text(28,19)[cb]{\small $32$}
\end{picture}
}
+
\scalebox{0.75}{
\begin{picture}(45,30)(-2,-2)
\Line(0,-15)(12,-6)
\Line(0,15)(12,6)
\Line(40,-15)(28,-6)
\Line(40,15)(28,6)
\CArc(20,0)(10,143,37)
\Line(12,15)(12,6)
\Text(12,12.5)[cb]{\scriptsize $\times$}
\Line(28,15)(28,6)
\Text(28,12.5)[cb]{\scriptsize $\times$}
\Text(-4, 12)[cb]{\small $2$}
\Text(-4,-18)[cb]{\small $8$}
\Text(45,-18)[cb]{\small $4$}
\Text(45, 12)[cb]{\small $1$}
\Text(12,19)[cb]{\small $16$}
\Text(28,19)[cb]{\small $32$}
\end{picture}
}
+
\scalebox{0.75}{
\begin{picture}(45,30)(-2,-2)
\Line(0,-15)(12,-6)
\Line(0,15)(12,6)
\Line(40,-15)(28,-6)
\Line(40,15)(28,6)
\CArc(20,0)(10,143,37)
\Line(12,15)(12,6)
\Text(12,12.5)[cb]{\scriptsize $\times$}
\Line(28,15)(28,6)
\Text(28,12.5)[cb]{\scriptsize $\times$}
\Text(-4, 12)[cb]{\small $8$}
\Text(-4,-18)[cb]{\small $4$}
\Text(45,-18)[cb]{\small $1$}
\Text(45, 12)[cb]{\small $2$}
\Text(12,19)[cb]{\small $16$}
\Text(28,19)[cb]{\small $32$}
\end{picture}
}
+
\scalebox{0.75}{
\begin{picture}(45,30)(-2,-2)
\Line(0,-15)(12,-6)
\Line(0,15)(12,6)
\Line(40,-15)(28,-6)
\Line(40,15)(28,6)
\CArc(20,0)(10,143,37)
\Line(12,15)(12,6)
\Text(12,12.5)[cb]{\scriptsize $\times$}
\Line(28,15)(28,6)
\Text(28,12.5)[cb]{\scriptsize $\times$}
\Text(-4, 12)[cb]{\small $4$}
\Text(-4,-18)[cb]{\small $1$}
\Text(45,-18)[cb]{\small $2$}
\Text(45, 12)[cb]{\small $8$}
\Text(12,19)[cb]{\small $16$}
\Text(28,19)[cb]{\small $32$}
\end{picture}
}
\nonumber\\[.4cm]
&&
{}+
\scalebox{0.75}{
\begin{picture}(45,30)(-2,-2)
\Line(0,-15)(12,-6)
\Line(0,15)(12,6)
\Line(40,-15)(28,-6)
\Line(40,15)(28,6)
\CArc(20,0)(10,143,37)
\Line(12,15)(12,6)
\Text(12,12.5)[cb]{\scriptsize $\times$}
\Line(28,15)(28,6)
\Text(28,12.5)[cb]{\scriptsize $\times$}
\Text(-4, 12)[cb]{\small $4$}
\Text(-4,-18)[cb]{\small $8$}
\Text(45,-18)[cb]{\small $2$}
\Text(45, 12)[cb]{\small $1$}
\Text(12,19)[cb]{\small $16$}
\Text(28,19)[cb]{\small $32$}
\end{picture}
}
+
\scalebox{0.75}{
\begin{picture}(45,30)(-2,-2)
\Line(0,-15)(12,-6)
\Line(0,15)(12,6)
\Line(40,-15)(28,-6)
\Line(40,15)(28,6)
\CArc(20,0)(10,143,37)
\Line(12,15)(12,6)
\Text(12,12.5)[cb]{\scriptsize $\times$}
\Line(28,15)(28,6)
\Text(28,12.5)[cb]{\scriptsize $\times$}
\Text(-4, 12)[cb]{\small $1$}
\Text(-4,-18)[cb]{\small $4$}
\Text(45,-18)[cb]{\small $8$}
\Text(45, 12)[cb]{\small $2$}
\Text(12,19)[cb]{\small $16$}
\Text(28,19)[cb]{\small $32$}
\end{picture}
}
+
\scalebox{0.75}{
\begin{picture}(45,30)(-2,-2)
\Line(0,-15)(12,-6)
\Line(0,15)(12,6)
\Line(40,-15)(28,-6)
\Line(40,15)(28,6)
\CArc(20,0)(10,143,37)
\Line(12,15)(12,6)
\Text(12,12.5)[cb]{\scriptsize $\times$}
\Line(28,15)(28,6)
\Text(28,12.5)[cb]{\scriptsize $\times$}
\Text(-4, 12)[cb]{\small $2$}
\Text(-4,-18)[cb]{\small $1$}
\Text(45,-18)[cb]{\small $4$}
\Text(45, 12)[cb]{\small $8$}
\Text(12,19)[cb]{\small $16$}
\Text(28,19)[cb]{\small $32$}
\end{picture}
}
+
\scalebox{0.75}{
\begin{picture}(45,30)(-2,-2)
\Line(0,-15)(12,-6)
\Line(0,15)(12,6)
\Line(40,-15)(28,-6)
\Line(40,15)(28,6)
\CArc(20,0)(10,143,37)
\Line(12,15)(12,6)
\Text(12,12.5)[cb]{\scriptsize $\times$}
\Line(28,15)(28,6)
\Text(28,12.5)[cb]{\scriptsize $\times$}
\Text(-4, 12)[cb]{\small $8$}
\Text(-4,-18)[cb]{\small $2$}
\Text(45,-18)[cb]{\small $1$}
\Text(45, 12)[cb]{\small $4$}
\Text(12,19)[cb]{\small $16$}
\Text(28,19)[cb]{\small $32$}
\end{picture}
}
\nonumber\\[.4cm]
&&
+
\scalebox{0.75}{
\begin{picture}(45,30)(-2,-2)
\Line(0,-15)(12,-6)
\Line(0,15)(12,6)
\Line(40,-15)(28,-6)
\Line(40,15)(28,6)
\CArc(20,0)(10,143,37)
\Line(12,15)(12,6)
\Text(12,12.5)[cb]{\scriptsize $\times$}
\Line(28,15)(28,6)
\Text(28,12.5)[cb]{\scriptsize $\times$}
\Text(-4, 12)[cb]{\small $1$}
\Text(-4,-18)[cb]{\small $8$}
\Text(45,-18)[cb]{\small $2$}
\Text(45, 12)[cb]{\small $4$}
\Text(12,19)[cb]{\small $16$}
\Text(28,19)[cb]{\small $32$}
\end{picture}
}
+
\scalebox{0.75}{
\begin{picture}(45,30)(-2,-2)
\Line(0,-15)(12,-6)
\Line(0,15)(12,6)
\Line(40,-15)(28,-6)
\Line(40,15)(28,6)
\CArc(20,0)(10,143,37)
\Line(12,15)(12,6)
\Text(12,12.5)[cb]{\scriptsize $\times$}
\Line(28,15)(28,6)
\Text(28,12.5)[cb]{\scriptsize $\times$}
\Text(-4, 12)[cb]{\small $8$}
\Text(-4,-18)[cb]{\small $2$}
\Text(45,-18)[cb]{\small $4$}
\Text(45, 12)[cb]{\small $1$}
\Text(12,19)[cb]{\small $16$}
\Text(28,19)[cb]{\small $32$}
\end{picture}
}
+
\scalebox{0.75}{
\begin{picture}(45,30)(-2,-2)
\Line(0,-15)(12,-6)
\Line(0,15)(12,6)
\Line(40,-15)(28,-6)
\Line(40,15)(28,6)
\CArc(20,0)(10,143,37)
\Line(12,15)(12,6)
\Text(12,12.5)[cb]{\scriptsize $\times$}
\Line(28,15)(28,6)
\Text(28,12.5)[cb]{\scriptsize $\times$}
\Text(-4, 12)[cb]{\small $2$}
\Text(-4,-18)[cb]{\small $4$}
\Text(45,-18)[cb]{\small $1$}
\Text(45, 12)[cb]{\small $8$}
\Text(12,19)[cb]{\small $16$}
\Text(28,19)[cb]{\small $32$}
\end{picture}
}
+
\scalebox{0.75}{
\begin{picture}(45,30)(-2,-2)
\Line(0,-15)(12,-6)
\Line(0,15)(12,6)
\Line(40,-15)(28,-6)
\Line(40,15)(28,6)
\CArc(20,0)(10,143,37)
\Line(12,15)(12,6)
\Text(12,12.5)[cb]{\scriptsize $\times$}
\Line(28,15)(28,6)
\Text(28,12.5)[cb]{\scriptsize $\times$}
\Text(-4, 12)[cb]{\small $4$}
\Text(-4,-18)[cb]{\small $1$}
\Text(45,-18)[cb]{\small $8$}
\Text(45, 12)[cb]{\small $2$}
\Text(12,19)[cb]{\small $16$}
\Text(28,19)[cb]{\small $32$}
\end{picture}
}
\nonumber\\[.4cm]
&&
{}+
\scalebox{0.75}{
\begin{picture}(45,30)(-2,-2)
\Line(0,-15)(12,-6)
\Line(0,15)(12,6)
\Line(40,-15)(28,-6)
\Line(40,15)(28,6)
\CArc(20,0)(10,143,37)
\Line(12,15)(12,6)
\Text(12,12.5)[cb]{\scriptsize $\times$}
\Line(28,15)(28,6)
\Text(28,12.5)[cb]{\scriptsize $\times$}
\Text(-4, 12)[cb]{\small $4$}
\Text(-4,-18)[cb]{\small $2$}
\Text(45,-18)[cb]{\small $8$}
\Text(45, 12)[cb]{\small $1$}
\Text(12,19)[cb]{\small $16$}
\Text(28,19)[cb]{\small $32$}
\end{picture}
}
+
\scalebox{0.75}{
\begin{picture}(45,30)(-2,-2)
\Line(0,-15)(12,-6)
\Line(0,15)(12,6)
\Line(40,-15)(28,-6)
\Line(40,15)(28,6)
\CArc(20,0)(10,143,37)
\Line(12,15)(12,6)
\Text(12,12.5)[cb]{\scriptsize $\times$}
\Line(28,15)(28,6)
\Text(28,12.5)[cb]{\scriptsize $\times$}
\Text(-4, 12)[cb]{\small $1$}
\Text(-4,-18)[cb]{\small $4$}
\Text(45,-18)[cb]{\small $2$}
\Text(45, 12)[cb]{\small $8$}
\Text(12,19)[cb]{\small $16$}
\Text(28,19)[cb]{\small $32$}
\end{picture}
}
+
\scalebox{0.75}{
\begin{picture}(45,30)(-2,-2)
\Line(0,-15)(12,-6)
\Line(0,15)(12,6)
\Line(40,-15)(28,-6)
\Line(40,15)(28,6)
\CArc(20,0)(10,143,37)
\Line(12,15)(12,6)
\Text(12,12.5)[cb]{\scriptsize $\times$}
\Line(28,15)(28,6)
\Text(28,12.5)[cb]{\scriptsize $\times$}
\Text(-4, 12)[cb]{\small $8$}
\Text(-4,-18)[cb]{\small $1$}
\Text(45,-18)[cb]{\small $4$}
\Text(45, 12)[cb]{\small $2$}
\Text(12,19)[cb]{\small $16$}
\Text(28,19)[cb]{\small $32$}
\end{picture}
}
+
\scalebox{0.75}{
\begin{picture}(45,30)(-2,-2)
\Line(0,-15)(12,-6)
\Line(0,15)(12,6)
\Line(40,-15)(28,-6)
\Line(40,15)(28,6)
\CArc(20,0)(10,143,37)
\Line(12,15)(12,6)
\Text(12,12.5)[cb]{\scriptsize $\times$}
\Line(28,15)(28,6)
\Text(28,12.5)[cb]{\scriptsize $\times$}
\Text(-4, 12)[cb]{\small $2$}
\Text(-4,-18)[cb]{\small $8$}
\Text(45,-18)[cb]{\small $1$}
\Text(45, 12)[cb]{\small $4$}
\Text(12,19)[cb]{\small $16$}
\Text(28,19)[cb]{\small $32$}
\end{picture}
}\;.
\label{without rules}
\\[-.2ex]\nonumber
\eeqar
The diagrams on the right-hand side are obtained by cutting in all
possible ways one of the loop lines of the diagrams on the left-hand
side.  The tree diagrams have been drawn in such a way, that one can
easily identify the original sequence of the loop lines (called ``loop
flow"), starting from the external loop leg with tag number $2^E$
and ending with the external loop leg with tag number $2^{E+1}$.
{Two simple rules can avoid the {redundant} tree diagrams and fix
  properly the correspondence with the loop diagrams:}
\begin{itemize}
\item [{1')}]
The {external} current $1$ must be attached to the external 
loop current with tag number $2^E$. 
This rule fixes the starting point of the loop flow and thus
reduces the redundancy already up to a factor of two, the
direction of the loop flow.
\item [{2')}] The {external} currents $1$, $2$ and $4$ must be
  attached to the loop flow in ascending order (the other {external
    currents} can enter the loop flow everywhere, also between $1$,
  $2$ and $4$). This rule uniquely fixes the direction of the loop
  flow.
\end{itemize}
With these rules the way of cutting each loop diagram of 
\refeq{without rules} becomes unique:
\beqar
\vcenter{\hbox{\scalebox{0.8}{
\begin{picture}(40,50)(0,-25)
\Line(0,0)(10,0)
\Line(40,-15)(28,-6)
\Line(40,15)(28,6)
\CArc(20,0)(10,0,360)
\Text(-4,-3)[cb]{\small $1$}
\Text(40,-24)[cb]{\small $2$}
\Text(40,18)[cb]{\small $4$}
\end{picture}}}
}
&\;\to\;&
\vcenter{\hbox{\scalebox{0.8}{
\begin{picture}(42,50)(0,-25)
\Line(0,0)(10,0)
\Line(40,-15)(28,-6)
\Line(40,15)(28,6)
\CArc(20,0)(10,143,37)
\Line(12,15)(12,6)
\Text(12,12.5)[cb]{\scriptsize $\times$}
\Line(28,15)(28,6)
\Text(28,12.5)[cb]{\scriptsize $\times$}
\Text(-4,-3)[cb]{\small $1$}
\Text(42,-24)[cb]{\small $2$}
\Text(42,18)[cb]{\small $4$}
\Text(12,19)[cb]{\small $8$}
\Text(28,19)[cb]{\small $16$}
\end{picture}}}
}\;,\nl[1.6ex]
\qquad
\vcenter{\hbox{\scalebox{0.75}{
\begin{picture}(40,50)(0,-25)
\Line(0,-15)(12,-6)
\Line(0,15)(12,6)
\Line(40,-15)(28,-6)
\Line(40,15)(28,6)
\CArc(20,0)(10,0,360)
\Text(0, 18)[cb]{\small $1$}
\Text(0,-24)[cb]{\small $2$}
\Text(40,-24)[cb]{\small $4$}
\Text(40, 18)[cb]{\small $8$}
\end{picture}}}
}
\;+\,
\vcenter{\hbox{\scalebox{0.75}{
\begin{picture}(40,50)(0,-25)
\Line(0,-15)(12,-6)
\Line(0,15)(12,6)
\Line(40,-15)(28,-6)
\Line(40,15)(28,6)
\CArc(20,0)(10,0,360)
\Text(0, 18)[cb]{\small $1$}
\Text(0,-24)[cb]{\small $2$}
\Text(40,-24)[cb]{\small $8$}
\Text(40, 18)[cb]{\small $4$}
\end{picture}}}
}
\;+\,
\vcenter{\hbox{\scalebox{0.75}{
\begin{picture}(40,50)(0,-25)
\Line(0,-15)(12,-6)
\Line(0,15)(12,6)
\Line(40,-15)(28,-6)
\Line(40,15)(28,6)
\CArc(20,0)(10,0,360)
\Text(0, 18)[cb]{\small $1$}
\Text(0,-24)[cb]{\small $8$}
\Text(40,-24)[cb]{\small $2$}
\Text(40, 18)[cb]{\small $4$}
\end{picture}}}
}
&\;\to\;&
\vcenter{\hbox{\scalebox{0.75}{
\begin{picture}(40,50)(0,-25)
\Line(0,-15)(12,-6)
\Line(0,15)(12,6)
\Line(40,-15)(28,-6)
\Line(40,15)(28,6)
\CArc(20,0)(10,143,37)
\Line(12,15)(12,6)
\Text(12,12.5)[cb]{\scriptsize $\times$}
\Line(28,15)(28,6)
\Text(28,12.5)[cb]{\scriptsize $\times$}
\Text(0, 18)[cb]{\small $1$}
\Text(0,-24)[cb]{\small $2$}
\Text(40,-24)[cb]{\small $4$}
\Text(40, 18)[cb]{\small $8$}
\Text(12,19)[cb]{\small $16$}
\Text(28,19)[cb]{\small $32$}
\end{picture}}}
}
\;+\,
\vcenter{\hbox{\scalebox{0.75}{
\begin{picture}(40,50)(0,-25)
\Line(0,-15)(12,-6)
\Line(0,15)(12,6)
\Line(40,-15)(28,-6)
\Line(40,15)(28,6)
\CArc(20,0)(10,143,37)
\Line(12,15)(12,6)
\Text(12,12.5)[cb]{\scriptsize $\times$}
\Line(28,15)(28,6)
\Text(28,12.5)[cb]{\scriptsize $\times$}
\Text(0, 18)[cb]{\small $1$}
\Text(0,-24)[cb]{\small $2$}
\Text(40,-24)[cb]{\small $8$}
\Text(40, 18)[cb]{\small $4$}
\Text(12,19)[cb]{\small $16$}
\Text(28,19)[cb]{\small $32$}
\end{picture}}}
}
\;+\,
\vcenter{\hbox{\scalebox{0.75}{
\begin{picture}(40,50)(0,-25)
\Line(0,-15)(12,-6)
\Line(0,15)(12,6)
\Line(40,-15)(28,-6)
\Line(40,15)(28,6)
\CArc(20,0)(10,143,37)
\Line(12,15)(12,6)
\Text(12,12.5)[cb]{\scriptsize $\times$}
\Line(28,15)(28,6)
\Text(28,12.5)[cb]{\scriptsize $\times$}
\Text(0, 18)[cb]{\small $1$}
\Text(0,-24)[cb]{\small $8$}
\Text(40,-24)[cb]{\small $2$}
\Text(40, 18)[cb]{\small $4$}
\Text(12,19)[cb]{\small $16$}
\Text(28,19)[cb]{\small $32$}
\end{picture}}}
}\;.
\label{rulesless}
\eeqar

The rules have to be generalised to other classes of diagrams
where external legs can combine in tree sub-graphs before entering
the loop flow.  To this end we define an identifier number for each
current, given by the smallest external tag among those forming its
tag number.  For example, a current with tag number $13=1+4+8$,
which has been created combining the external legs $1$, $4$ and $8$,
has identifier $1$; a current with tag number $6=2+4$ has
identifier $2$.  For external currents the identifier
coincides with the tag number.  In case of a quadri-linear coupling,
the two currents entering the loop are represented by a common
identifier, the minimum of the two identifiers for the individual
currents.

Now the generalisation of the two rules is straightforward:
\begin{itemize}
\item [1)] The current with identifier $1$ must be attached to
  the external loop current with tag number $2^E$.
\item [2)] The currents with the three smallest identifiers must be
  attached to the loop flow following the ascending order of
  {their} identifiers.
\end{itemize}

The proper treatment of self-energy insertions deserves particular
care.  For two identical particles flowing in the self-energy loop,
the selection rules (actually the first rule alone) reduce the number
of cut diagrams to one.  On the other hand in this case the
self-energy diagram gets a symmetry factor $1/2$ because of Wick's
theorem.  If the two particles in the loop are different, we get two
cut diagrams, which however give the same contribution.  Therefore,
the cutting procedure for self-energies reproduces the loop diagrams
correctly if the corresponding tree contributions are multiplied by a
factor $1/2$ in all cases.

In addition to rules 1) and 2), in most renormalisation schemes some
classes of diagrams have to be discarded, namely {tadpoles and
  self-energy insertions on external legs.}  Therefore, {being $S = 1 +
2 + 4 + \dots + 2^{E-1}{=2^E-1}$} the sum of all external tags of the
process, we {use} the following additional rules:
\begin{itemize}
\item [3)] A current with tag number equal to $S$ or equal to $S-2^n$,
  $n=0,1,\dots E-1$, cannot enter the loop flow in a branch with a
  tri-linear vertex.  This eliminates tadpole diagrams and those
  self-energy contributions made of tri-linear vertices which are
  inserted on external legs.
\item [4)] {If in a branch with a quadri-linear vertex} one of the 
  two currents entering the loop flow is external, the sum of their 
  tag numbers cannot be equal to $S$.  This eliminates {the self-energy 
  contributions involving} one quadri-linear vertex inserted on external legs.
\end{itemize}
Applying these four rules to the generation of the skeleton for the
off-shell currents of the tree processes of the set $\{A\to
B+P+\bar{P},\,\forall P\in\SM\}$, we obtain the proper skeleton for
the one-loop process $A\to B$.

Having reduced the formal generation of the one-loop amplitude to the
generation of a set of tree-level processes, we can build the ``loop
off-shell currents" in a similar way as the tree-level currents in
\refse{tree}, in order to obtain the tensor coefficients
{$c_{\mu_1\cdots\mu_{r_j}}^{(j,r_j,N_j)}$} of \refeq{tensor
  splitting}.
At one-loop level, the particle which closes the recursive
construction is the external loop leg with tag number $2^{E+1}${.
  Since} all loop lines are virtual lines and {retain} their
propagator, the last step of \refeq{last step}, where the last
generated current is multiplied with the wave function of the particle
closing the recursion, $\begin{picture}(53,10)(-5,-2) \Line(0,0)(20,0)
  \GCirc(0,0){1.5}{0} \Text(20,-3)[cb]{\scriptsize $\times$}
  \Text(37,-3)[cb]{$2^{E+1}$}
\end{picture}$,
is performed without multiplication by the inverse propagator
resulting in
\vspace{-10pt}
\beq
{\de\M=}
\vcenter{\hbox{
\begin{picture}(60,70)(0,-35)
\Line(5,-23)(17,-6)
\Line(35,-23)(23,-6)
\CArc(20,0)(10,-45,225)
\GOval(20,-7)(10,3)(90){.8}
\DashCArc(20,0)(22,250,290){2}
\Text(5,-34)[cb]{\small $1$}
\Text(40,-34)[cb]{\small $2^{E-1}$}
\end{picture}}}
{\leftrightarrow\;\;\sum_P}\vcenter{\hbox{
\begin{picture}(55,60)(0,-30)
\Line(5,-23)(17,-6)
\Line(35,-23)(23,-6)
\Line(12,-6)(12,8)
\Text(12,5.5)[cb]{\scriptsize $\times$}
\Line(28,-6)(28,8)
\GCirc(28,8){1.5}{0}
\GOval(20,-7)(10,3)(90){.8}
\DashCArc(20,0)(22,250,290){2}
\Text(5,-34)[cb]{\small $1$}
\Text(40,-34)[cb]{\small $2^{E-1}$}
\Text(15,14)[cb]{$2^E$} 
\Text(35,-3)[cb]{\small $P$}
\end{picture}}}
\times\vcenter{\hbox{
\begin{picture}(30,10)(-5,-5)
\Line(0,0)(20,0)
\GCirc(0,0){1.5}{0}
\Text(20,-3)[cb]{\scriptsize $\times$}
\Text(10,-10)[cb]{\small $P$}
\Text(28,-3)[lb]{$2^{E+1}\;.$} 
\end{picture}}}
\eeq
The external currents for the first $E$ {external legs} are defined
as in \refse{tree}; the external currents of the two external loop
legs are defined such that the contraction originally contained in
the loop can be easily reproduced.  To this end, we introduce a
suitable set of spinors $\psi_i=u_i,{\bar{v}}_i$ and polarisation
vectors {$\epsilon_i^\mu$} for the cut fermions and vector bosons,
\beqar
(\psi_i)_{\alpha} = (\bar{\psi}_i)_{\alpha} = \delta_{i\alpha},
\qquad
&\mbox{with}&\quad
\sum_{i=1}^4(\bar{\psi}_i)_{\alpha}(\psi_i)_{\beta} = \delta_{\alpha\beta},\nl
\epsilon_i^\mu = \delta_i^\mu,
\qquad&
\mbox{with}&\quad
\sum_{i=1}^4\epsilon_i^\mu\epsilon_i^\nu = \delta^{\mu\nu},
\label{cutspinors}
\eeqar
where $i$ denotes the ``polarisation'', and $\alpha,\beta$ and
$\mu,\nu$ are spinor and Lorentz indices, respectively.  The loops are
glued together as:

\vspace{-1cm}
\beqar
\mbox{scalars:}
&&
\qquad
\begin{picture}(25,40)(6,-8)
\Line(5,-23)(17,-6)
\Line(35,-23)(23,-6)
\CArc(20,0)(10,-45,225)
\GOval(20,-7)(10,3)(90){.8}
\DashCArc(20,0)(22,250,290){2}
\end{picture}
\quad\,\leftrightarrow\quad
\qquad\!\quad
\begin{picture}(25,40)(6,-8)
\Line(5,-23)(17,-6)
\Line(35,-23)(23,-6)
\Line(12,-6)(12,8)
\Text(12,5.5)[cb]{\scriptsize $\times$}
\Line(28,-6)(28,8)
\GCirc(28,8){1.5}{0}
\GOval(20,-7)(10,3)(90){.8}
\DashCArc(20,0)(22,250,290){2}
\Text(13,13)[cb]{\small $1$}
\end{picture}
\quad\times\quad
\begin{picture}(25,10)(-2,-2)
\Line(0,0)(18,0)
\GCirc(0,0){1.5}{0}
\Text(18,-3)[cb]{\scriptsize $\times$}
\Text(25,-3)[cb]{\small $1\,,$}
\end{picture}
\qquad\qquad
\nonumber\\[.3cm]
\mbox{vector bosons:}
&&
\qquad
\begin{picture}(25,40)(6,-8)
\Line(5,-23)(17,-6)
\Line(35,-23)(23,-6)
\PhotonArc(20,0)(10,-45,225){1.5}{10}
\GOval(20,-7)(10,3)(90){.8}
\DashCArc(20,0)(22,250,290){2}
\end{picture}
\quad\,\leftrightarrow\,\quad
\sum_{i=1}^4\,\quad
\begin{picture}(25,40)(6,-8)
\Line(5,-23)(17,-6)
\Line(35,-23)(23,-6)
\Photon(12,-6)(12,8){1.5}{4.5}
\Text(12,5.5)[cb]{\scriptsize $\times$}
\Photon(28,-6)(28,8){1.5}{4.5}
\GCirc(28,8){1.5}{0}
\GOval(20,-7)(10,3)(90){.8}
\DashCArc(20,0)(22,250,290){2}
\Text(13,13)[cb]{\small $\epsilon_i$}
\end{picture}
\quad\times\quad
\begin{picture}(25,10)(-2,-2)
\Photon(0,0)(18,0){1.5}{5}
\GCirc(0,0){1.5}{0}
\Text(18,-3)[cb]{\scriptsize $\times$}
\Text(27,-3)[cb]{\small $\epsilon_i\,,$}
\end{picture}
\nonumber\\
\mbox{fermions:}
&&
\qquad
\begin{picture}(25,40)(6,-8)
\Line(5,-23)(17,-6)
\Line(35,-23)(23,-6)
\ArrowArc(20,0)(10,-45,225)
\GOval(20,-7)(10,3)(90){.8}
\DashCArc(20,0)(22,250,290){2}
\end{picture}
\quad\,\leftrightarrow\quad
\sum_{i=1}^4\,\quad
\begin{picture}(25,40)(6,-8)
\Line(5,-23)(17,-6)
\Line(35,-23)(23,-6)
\ArrowLine(12,8)(12,-6)
\Text(12,5.5)[cb]{\scriptsize $\times$}
\ArrowLine(28,-6)(28,8)
\GCirc(28,8){1.5}{0}
\GOval(20,-7)(10,3)(90){.8}
\DashCArc(20,0)(22,250,290){2}
\Text(13,13)[cb]{\small $\psi_i$}
\end{picture}
\quad\times\quad
\begin{picture}(25,10)(-2,-2)
\ArrowLine(0,0)(18,0)
\GCirc(0,0){1.5}{0}
\Text(18,-3)[cb]{\scriptsize $\times$}
\Text(29,-3)[cb]{\small $\bar{\psi}_i\,,$}
\end{picture}
\nonumber\\
&&
\qquad
\begin{picture}(25,40)(6,-8)
\Line(5,-23)(17,-6)
\Line(35,-23)(23,-6)
\ArrowArcn(20,0)(10,225,-45)
\GOval(20,-7)(10,3)(90){.8}
\DashCArc(20,0)(22,250,290){2}
\end{picture}
\quad\,{\leftrightarrow}\quad
\sum_{i=1}^4\,\quad
\begin{picture}(25,40)(6,-8)
\Line(5,-23)(17,-6)
\Line(35,-23)(23,-6)
\ArrowLine(12,-6)(12,8)
\Text(12,5.5)[cb]{\scriptsize $\times$}
\ArrowLine(28,8)(28,-6)
\GCirc(28,8){1.5}{0}
\GOval(20,-7)(10,3)(90){.8}
\DashCArc(20,0)(22,250,290){2}
\Text(13,13)[cb]{\small $\bar{\psi}_i$}
\end{picture}
\quad\times\quad
\begin{picture}(25,10)(-2,-2)
\ArrowLine(18,0)(0,0)
\GCirc(0,0){1.5}{0}
\Text(18,-3)[cb]{\scriptsize $\times$}
\Text(29,-3)[cb]{\small $\psi_i\,.$}
\end{picture} 
\label{cut}
\eeqar
Except for the scalar case, the cutting procedure associates to the
one-loop amplitude the sum of four tree-level amplitudes with {particular}
spinors/polarisation vectors for the cut particle.

Having fixed the external currents, we describe how to compute the
internal ones.  As explained in \refse{tree}, for tree amplitudes
these are computed {summing up the currents generated in} branches
where the generating currents are multiplied with the Feynman rules
for the vertex and the propagator of the generated particle.  This is
valid also at one-loop level for pure tree currents built by combining
the original external legs {$1,\ldots,2^{E-1}$}.

The new features of the loop case are connected to the loop
{off-shell} currents involving the external loop leg with tag
number $2^E$ {carrying} a loop momentum $q$.  The external {loop} current
with tag number $2^E$ defines the beginning of the loop flow; all
currents with tag number $\ge 2^E$ belong to the loop flow and are
called loop currents, while the branches generating them are called
loop branches.  Every internal loop current contains a $q$-dependence,
generated by the Feynman rules for the vertex and the propagator.
Working in the 't Hooft--Feynman gauge in the SM, the $q$-dependence
of (vertex)$\times$(propagator) takes the form
\beq
(\mbox{vertex}) \;\times\; (\mbox{propagator}) = 
\frac{a^\mu q_\mu + b}{(q+p)^2-m^2},
\label{loop vertexprop}
\eeq
where the linear $q$-dependence in the numerator results from a
fermion propagator, from a coupling of three vector bosons, or from a
coupling between one vector boson and two scalar{{/ghost}} particles
(in all other cases $a^\mu=0$).  If the interacting particles are
fermions and/or vector bosons, the coefficients $a^\mu$ and $b$ have
an additional Dirac {and/or} Lorentz structure which {is not
  made explicit} here for simplicity.
{Denoting} by $w_1(q)$ the first internal loop current, 
we have
\beq
w_1(q) =
\frac{d_{1,1} ^{\mu_1} q_{\mu_1} + d_{1,0}}{(q+p_1)^2-m_1^2},
\label{w1}
\eeq
where $p_1$ is the sum of the external momenta {entering} the first
loop branch while $d_{1,1}^{\mu_1} $ and $d_{1,0}$ result from {a}
product of the generating currents with the constants $a^\mu$ and $b$
in \refeq{loop vertexprop}.  If the current $w_1$ corresponds to a
fermion/vector boson, {$w_1$} as well as $d_{1,1}^{\mu_1}$ and
$d_{1,0}$ carry an additional spinor/Lorentz index, {suppressed} in
(\ref{w1}).  Proceeding along the loop flow, the second internal loop
current is built combining $w_1(q)$ with tree currents and with a
product (vertex)$\times$(propagator) of the form~\refeq{loop
  vertexprop} and reads
\beq
w_2(q) =
\frac{d_{2,2}^{\mu_1\mu_2}q_{\mu_1}q_{\mu_2} + d_{2,1}^{\mu_1} q_{\mu_1} + d_{2,0}}
     {[(q+p_1)^2-m_1^2][(q+p_2)^2-m_2^2]}.
\eeq
The {recursively constructed} $l^{\rm th}$ loop current of the
loop flow is of the form
\beq
w_l(q) =
\sum_{k=0}^l
\frac{d_{l,k}^{\mu_1\cdots\mu_k}q_{\mu_1}\cdots q_{\mu_k}}
     {\prod_{h={1}}^l[(q+p_h)^2-m_h^2]}.
\label{wl}
\eeq
For the last loop current (with $l ={N_j}=$ number of loop lines)
the momentum $p_{{N_j}}$ is equal to the sum of all external
momenta and thus vanishes.

The recursion relations {\refeq{recursive}} are valid also for the loop
currents, but cannot be used to compute them numerically (unless we
give an explicit value to the loop momentum $q$).  However, one can
define similar relations to compute the set of coefficients
$\{d_{l,0},d_{l,1}^{\mu_1},\dots,d_{l,l}^{\mu_1\cdots\mu_l}\}$, using
the general form of \refeq{loop vertexprop} for the $q$-dependence of
loop branches.  In fact in a {loop branch with a tri-linear vertex,}
knowing the generating tree current $w_t$ and the coefficients
$\{d_{l-1,0},d_{l-1,1}^{\mu_1},\dots,d_{l-1,l-1}^{\mu_1\cdots\mu_{l-1}}\}$
of the generating loop current, the coefficients
$\{d_{l,0},d_{l,1}^{\mu_1},\dots,d_{l,l}^{\mu_1\cdots\mu_l}\}$ of the
generated {loop} current are given by
\beqar
\{d_{l,0},d_{l,1}^{\mu_1},\dots,d_{l,l}^{\mu_1\cdots\mu_l}\}
&=&
w_t\,\bigg(
\{0,d_{l-1,0},d_{l-1,1}^{\mu_2},\dots,d_{l-1,l-1}^{\mu_2\cdots\mu_l}\}\,a^{\mu_1} 
\nonumber\\
&&\qquad
{}+ \,
\{d_{l-1,0},d_{l-1,1}^{\mu_1},\dots,d_{l-1,l-1}^{\mu_1\cdots\mu_{l-1}},0\}\,b
\;\bigg),
\label{recursion loop}
\eeqar
where we have again omitted the Dirac/Lorentz indices {associated
  with fermionic or vectorial currents} as in \refeq{w1}.  For
{quadri-linear} vertices, the situation is even simpler because in
this case $a^\mu=0$ and \refeq{recursion loop} simplifies to
\beq
\{d_{l,0},d_{l,1}^{\mu_1},\dots,d_{l,l}^{\mu_1\cdots\mu_l}\}
\,=\,
w_{t_1}w_{t_2}\,
\{d_{l-1,0},d_{l-1,1}^{\mu_1},\dots,d_{l-1,l-1}^{\mu_1\cdots\mu_{l-1}},0\}\,b,
\label{recursion loop 4}
\eeq
where $w_{t_1}$ and $w_{t_2}$ are the two generating tree-level
currents.  The expressions \refeq{recursion loop} and \refeq{recursion
  loop 4} have to be used at each loop branch.  The {generated}
coefficients $d_{l,k}^{\mu_1\cdots\mu_k}$, $k=0,\dots,l$ are in
general not symmetric under the exchange of their indices
$\mu_1\cdots\mu_k$, but, being implicitly multiplied by the symmetric
product $q_{\mu_1}\cdots q_{\mu_k}$, they can be symmetrised at each
step.  In this way the number of independent coefficients
$d_{l,k}^{\mu_1\cdots\mu_k}$ is {decreased,} leading to a reduction of
operations in subsequent steps of the recursion.

The recursion relations for loop branches allow us to compute the
coefficients of the loop currents, but we are not allowed to sum them
unless their denominators are equal.  From \refeq{wl} one can see that
the denominators are products of propagators and are determined by a
sequence of off-set momenta $\{p_1,\dots,p_l\}$ and masses
$\{m_1,\dots,m_l\}$.  Therefore, while tree currents are defined by
the tag number, the particle content, and the colour information, the
loop currents need an additional parameter, called sequence number,
containing the information on $\{p_1,\dots,p_l\}$ and
$\{m_1,\dots,m_l\}$.  In this way, loop currents with different
$q$-dependent denominators are distinguished, and contributions from
loop branches with the same denominators can be summed as for tree
branches.

The introduction of the sequence number spoils the uniqueness of the
last current.  Given the cut particle and its polarisation [the index
$i$ {in} \refeq{cutspinors}], the coefficients
$\{d_{N_j,0},d_{N_j,1}^{\mu_1},\dots,d_{N_j,
  N_j}^{\mu_1\cdots\mu_{N_j}}\}$ of the last current with sequence
number $n_s$ give a contribution to the tensor coefficients of
\refeq{tensor splitting} with $j = n_s$.  The index $j$, describing
the class of the tensor integrals, is then identified with the
sequence number of the last currents.  At this level, contributions
from different polarisations of the cut particle can be summed up.
For different cut particles one gets in general contributions to
different {classes of tensor integrals.} However, since {not the
  particles but only their masses and momenta} enter the sequence
number, also contributions to the same tensor integral classes
appear and are combined.

Also at one-loop level the code is divided into an initialisation and
a production phase. {In} the initialisation phase, the skeleton of the
branches is generated and all quantities which do not depend on the
momenta are fixed.  In particular, the sequence numbers of the last
currents allow already at this step to determine the list of needed
tensor integrals. Therefore the computation of the tensor integrals
can be done independently {of} the one of the tensor coefficients (in
the production phase of the code).
\subsection {Rational terms and renormalisation}
\label{R2&CT}
In dimensional regularisation the calculation of Feynman amplitudes is
performed in $D=4-2\epsilon$ space-time dimensions, and the result is
arranged as a power series in $\epsilon$. In this way, UV divergences
of the loop integrals manifest themselves as $1/\epsilon$ poles to be
subtracted {upon} renormalisation. For consistency, all space-time
related objects entering the amplitude have to be promoted to their
$D$-dimensional generalisations and all manipulations have to be
performed in $D$ dimensions. Otherwise terms of order
$\mathcal{O}(\epsilon)$ are missed which, combined with the
$1/\epsilon$ pole of the loop integral, give finite contributions to
the amplitude. Because these are rational functions of the kinematical
invariants, they are conventionally called rational terms. A rational
term is dubbed $R_1$-term if it results from the $\epsilon$-dependence
of the denominators of the loop integrals and it is called $R_2$-term
if it is generated by a $\mathcal{O}(\epsilon)$ term in the numerator
of the Feynman amplitude~\cite{Ossola:2008xq}.

We assume that the tensor integrals, taken as input by \recola,
contain the $R_1$-terms, either by keeping the denominators of the
loop propagators $D$-dimensional in the calculation or by explicitly
adding these terms. Note, however, that even {if} the calculation of
the tensor integrals is performed in $D$ dimensions, they enter
\recola\ as numerical four-dimensional tensors.  The numerical
construction of the tensor coefficients, on the other hand, works
strictly in four space-time dimensions, so that the $R_2$-terms are
not taken into account automatically.  These terms {can, however, be
  easily} computed using effective Feynman rules which have been
implemented in our code using the results of
\citeres{Ossola:2008xq,Draggiotis:2009yb,Garzelli:2009is,Shao:2011tg}\,\footnote{We thank R.~Pittau for clarifications concerning
  \citeres{Garzelli:2009is,Shao:2011tg}.}.
The insertion of the effective {Feynman} rules for vertices and
  propagators is performed in the tree-level amplitude generator,
  {taking care} that only one of the vertices results from a
rational term.
  
Renormalisation is performed via counterterms based on the conventions
of \citere{Denner:1991kt}. In analogy with the effective Feynman rules
for the rational terms, the insertion of counterterms takes place in
the tree-level amplitude generator.  Presently, counterterms are fixed
following the complex-mass scheme of
\citeres{Denner:1999gp,Denner:2005fg} and the results of
\citere{Denner:1991kt} for all parameters of the SM. The strong
coupling constant is renormalised in the
$\overline{\mathrm{MS}}$-scheme at a general scale $Q$ for
contributions coming from gluons and light quarks, while the top-quark
contribution is subtracted at zero momentum.

\subsection{Treatment of colour}
\label{colour}
In the computation of the currents an important aspect is the
treatment of colour, which does not factorise in the recursive
construction (contrary to the diagrammatic approach).  One could
obtain the factorisation of colour by splitting the amplitude in a sum
of colour-ordered amplitudes, as done in \citere{Bern:1990ux}. This,
however, would increase the number of amplitudes to compute and
{would} become complicated in the full SM.  Alternatively, one
could compute colour-dressed amplitudes, as for instance in
\citere{Kanaki:2000ey}, where off-shell currents would carry
{explicit} colour indices and would have to be computed for each index
separately, slowing down the calculation considerably.  Although the
number of {colour-dressed amplitudes to be computed} can be
{decreased} by a Monte Carlo sampling over colour
configurations\cite{Papadopoulos:2005ky}, the number of operations at
intermediate steps remains large.  In order to optimise the colour
treatment further, we developed an alternative approach based on
``structure-dressed" amplitudes, where each current gets an explicit
colour structure.

This is easily achieved working in the colour-flow representation of
the {$1 \slash \NC$} expansion~\cite{'tHooft:1973jz}, introduced
in \citeres{Kanaki:2000ms,Maltoni:2002mq} for perturbative QCD
computations, where the conventional 8 gluon fields $A_{\mu}^{a}$ {are
  replaced by} a $3\times 3$ matrix $({\cal A}_{\mu})^{i}_{j} =
\frac{1}{\sqrt{2}}\,A_{\mu}^{a}(\lambda^{a})^{i}_{j}$ with the trace
condition $\sum_{i}\,({\cal A}_{\mu})^{i}_{i} = 0$.  Quarks and
antiquarks maintain the usual colour index $i=1,{2},3$, while gluons
get a pair of indices $i,j=1,{2},3$; the Gell-Mann matrices
{$\lambda^{a}$} and the structure constants in the Feynman rules
are then substituted by combinations of Kronecker $\delta$s.  The
propagators read
\beqar
\begin{picture}(40,10)(-5,-2)
\ArrowLine(0,0)(25,0)
\Text(-4,-4)[cb]{{$j$}}
\Text(30,-2)[cb]{{$i$}}
\Text(13,-12)[cb]{$p$}
\end{picture}
&=&
\delta^{i}_{j}
\times
\frac{\ri(\ps+m)}{p^2-m^2},
\nonumber\\
\begin{picture}(65,10)(-20,-2)
\Gluon(0,0)(25,0){2.5}{4}
\Text(-6,1)[cb]{$i_1$}
\Text(-6,-9)[cb]{$j_1$}
\Text(32,-9)[cb]{$i_2$}
\Text(32,1)[cb]{$j_2$}
\Text(-18,-3)[cb]{$\mu$}
\Text(44,-3)[cb]{$\nu$}
\Text(13,-12)[cb]{$p$}
\end{picture}
&=&
\begin{picture}(50,10)(-12,-2)
\ArrowLine(25,3)(0,3)
\ArrowLine(0,-3)(25,-3)
\Text(-6,1)[cb]{$i_1$}
\Text(-6,-9)[cb]{$j_1$}
\Text(32,-9)[cb]{$i_2$}
\Text(32,1)[cb]{$j_2$}
\end{picture}
\times
\frac{-\,\ri\,g_{\mu\nu}}{p^2}
\;
=
\;
\delta^{i_1}_{j_2}\delta^{i_2}_{j_1}\,
\frac{-\,\ri\,g_{\mu\nu}}{p^2},
\nonumber\\
\begin{picture}(65,10)(-20,-2)
\DashArrowLine(0,0)(25,0){3}
\Text(-6,1)[cb]{$i_1$}
\Text(-6,-9)[cb]{$j_1$}
\Text(32,-9)[cb]{$i_2$}
\Text(32,1)[cb]{$j_2$}
\Text(13,-12)[cb]{$p$}
\end{picture}
&=&
\begin{picture}(50,10)(-12,-2)
\ArrowLine(25,3)(0,3)
\ArrowLine(0,-3)(25,-3)
\Text(-6,1)[cb]{$i_1$}
\Text(-6,-9)[cb]{$j_1$}
\Text(32,-9)[cb]{$i_2$}
\Text(32,1)[cb]{$j_2$}
\end{picture}
\times
\frac{\ri}{p^2}
\;
=
\;
\delta^{i_1}_{j_2}\delta^{i_2}_{j_1}\,
\frac{\ri}{p^2},
\eeqar
{while the vertices become}
\vspace{-.7cm}
\beqar
\begin{picture}(75,50)(-8,-2)
\Gluon(20,0)(43,0){2.5}{4}
\ArrowLine(20,0)(0,20)            
\ArrowLine(0,-20)(20,0)
\Vertex(20,0){2}
\Text(-5,19)[cb]{$i_1$} 
\Text(-5,-26)[cb]{$j_2$}
\Text(52,-9)[cb]{$i_3$}
\Text(52,1)[cb]{$j_3$}
\Text(64,-3)[cb]{$\mu$}
\end{picture}
&=&
\,
\left(
\scalebox{0.8}{
\begin{picture}(60,35)(-8,-2)
\ArrowLine(17,3)(0,20)     
\ArrowLine(40,3)(17,3)
\ArrowLine(0,-20)(17,-3)
\ArrowLine(17,-3)(40,-3)
\Text(-5,19)[cb]{$i_1$} 
\Text(-5,-24)[cb]{$j_2$}
\Text(50,-9)[cb]{$i_3$}
\Text(50,1)[cb]{$j_3$}
\end{picture}
}
 - \;\frac{1}{\NC}
\scalebox{0.8}{
\begin{picture}(60,35)(-8,-2)
\ArrowLine(17,3)(0,20)
\ArrowLine(0,-20)(17,-3)
\CArc(14,0)(4.2,-45,45)
\DashLine(18.2,0)(30.8,0){3}
\CArc(35,0)(3,90,270)
\ArrowLine(43,3)(35,3)
\ArrowLine(35,-3)(43,-3)
\Text(-5,19)[cb]{$i_1$} 
\Text(-5,-24)[cb]{$j_2$}
\Text(52,-9)[cb]{$i_3$}
\Text(52,1)[cb]{$j_3$}
\end{picture}
}
\right)
\times
\frac{\ri\,\gs}{\sqrt{2}}\,\gamma^\mu
=
\;
\left(
\delta^{i_1}_{j_3}\delta^{i_3}_{j_2}
- \frac{1}{\NC}\delta^{i_1}_{j_2}\delta^{i_3}_{j_3}
\right)
\frac{\ri\,\gs}{\sqrt{2}}\,\gamma^\mu,
\nonumber\\[.4cm]
\begin{picture}(80,40)(-16,-2)
\Gluon(20,0)(43,0){2.5}{4}
\Gluon(20,0)(0,20){2.5}{5}
\Gluon(0,-20)(20,0){2.5}{5}
\Vertex(20,0){2}
\Text(1,25)[cb]{$i_1$} 
\Text(-6,18)[cb]{$j_1$} 
\Text(-7,-24)[cb]{$i_2$}
\Text(2,-31)[cb]{$j_2$}
\Text(50,-9)[cb]{$i_3$}
\Text(50,1)[cb]{$j_3$}
\Text(-13,29)[cb]{$\mu$}
\Text(-13,-34)[cb]{$\nu$}
\Text(62,-3)[cb]{$\rho$}
\LongArrow(10,20)(20,10)\Text(24,15)[cb]{$p_1$}
\LongArrow(-2,-12)(8,-2)\Text(-3,-5)[cb]{$p_2$}
\LongArrow(38,-7)(25,-7)\Text(33,-17)[cb]{$p_3$}
\end{picture}
&=&
\left(
\scalebox{0.75}{
\begin{picture}(60,35)(-10,-2)
\ArrowLine(17,3)(0,20)     
\ArrowLine(40,3)(17,3)
\ArrowLine(-5,15)(12,0)
\ArrowLine(12,0)(-5,-15)
\ArrowLine(0,-20)(17,-3)
\ArrowLine(17,-3)(40,-3)
\Text(-2,22)[cb]{$i_1$} 
\Text(-9,15)[cb]{$j_1$} 
\Text(-9,-20)[cb]{$i_2$}
\Text(-2,-27)[cb]{$j_2$}
\Text(48,-9)[cb]{$i_3$}
\Text(48,1)[cb]{$j_3$}
\end{picture}
}
-
\scalebox{0.75}{
\begin{picture}(60,35)(-10,-2)
\ArrowLine(0,20)(17,3)
\ArrowLine(17,3)(40,3)
\ArrowLine(12,0)(-5,15)
\ArrowLine(-5,-15)(12,0)
\ArrowLine(17,-3)(0,-20)
\ArrowLine(40,-3)(17,-3)
\Text(-2,22)[cb]{$j_1$} 
\Text(-9,15)[cb]{$i_1$} 
\Text(-9,-20)[cb]{$j_2$}
\Text(-2,-27)[cb]{$i_2$}
\Text(48,-9)[cb]{$j_3$}
\Text(48,1)[cb]{$i_3$}
\end{picture}
}
\right)
\frac{\ri\,\gs}{\sqrt{2}}\,\Big[
  g^{\mu\nu}\!(p_1\!-\!p_2)^{\rho}\!
+ \!g^{\nu\!\rho}(p_2\!-\!p_3)^{\mu}\!
+ \!g^{\rho\mu}(p_3\!-\!p_1)^{\!\nu}
\Big]
\nonumber\\
&=&
\left(
{
  \delta^{i_1}_{j_3}\delta^{i_2}_{j_1}\delta^{i_3}_{j_2}
- \delta^{i_1}_{j_2}\delta^{i_2}_{j_3}\delta^{i_3}_{j_1}
}
\right)
\frac{\ri\,\gs}{\sqrt{2}}\,\Big[
  g^{\mu\nu}(p_1\!-\!p_2)^{\rho}\!
+ \!g^{\nu\rho}(p_2\!-\!p_3)^{\mu}\!
+ \!g^{\rho\mu}(p_3\!-\!p_1)^{\nu}
\Big],
\nonumber\\[.4cm]
\begin{picture}(80,40)(-16,-2)
\Gluon(0,20)(20,0){2.5}{5}
\Gluon(0,-20)(20,0){2.5}{5}
\Gluon(40,-20)(20,0){2.5}{5}
\Gluon(40,20)(20,0){2.5}{5}
\Vertex(20,0){2}
\Text(1,25)[cb]{$i_1$} 
\Text(-6,18)[cb]{$j_1$} 
\Text(-7,-24)[cb]{$i_2$}
\Text(2,-31)[cb]{$j_2$}
\Text(40,-31)[cb]{$i_3$}
\Text(47,-24)[cb]{$j_3$}
\Text(49,16)[cb]{$i_4$} 
\Text(40,24)[cb]{$j_4$} 
\Text(-13,29)[cb]{$\mu$}
\Text(-13,-34)[cb]{$\nu$}
\Text(53,-34)[cb]{$\rho$}
\Text(53,29)[cb]{$\si$}
\end{picture}
&=&
\frac{\ri\,\gs^2}{2}\,\Big[
\quad
\Big(
  \delta^{i_1}_{j_4}\delta^{i_2}_{j_1}\delta^{i_3}_{j_2}\delta^{i_4}_{j_3}
+ \delta^{i_1}_{j_2}\delta^{i_2}_{j_3}\delta^{i_3}_{j_4}\delta^{i_4}_{j_1}
\Big)
\Big(
  2g^{\mu\rho}g^{\nu\si}
- g^{\mu\si}g^{\nu\rho}
- g^{\mu\nu}g^{\rho\si}
\Big)
\nonumber\\
&&
\qquad\;+\,
\Big(
  \delta^{i_1}_{j_3}\delta^{i_2}_{j_4}\delta^{i_3}_{j_2}\delta^{i_4}_{j_1}
+ \delta^{i_1}_{j_4}\delta^{i_2}_{j_3}\delta^{i_3}_{j_1}\delta^{i_4}_{j_2}
\Big)
\Big(
 2g^{\mu\nu}g^{\rho\si}
- g^{\mu\rho}g^{\nu\si}
- g^{\mu\si}g^{\nu\rho}
\Big)
\nonumber\\
&&
\qquad\;+\,
\Big(
  \delta^{i_1}_{j_3}\delta^{i_2}_{j_1}\delta^{i_3}_{j_4}\delta^{i_4}_{j_2}
+ \delta^{i_1}_{j_2}\delta^{i_2}_{j_4}\delta^{i_3}_{j_1}\delta^{i_4}_{j_3}
\Big)
\Big(
 2g^{\mu\si}g^{\nu\rho}
- g^{\mu\nu}g^{\rho\si}
- g^{\mu\rho}g^{\nu\si}
\Big)
\Big],
\nonumber\\
\begin{picture}(80,40)(-16,-2)
\Gluon(20,0)(43,0){2.5}{4}
\DashArrowLine(20,0)(0,20){3}
\DashArrowLine(0,-20)(20,0){3}
\Vertex(20,0){2}
\Text(1,25)[cb]{$i_1$} 
\Text(-6,18)[cb]{$j_1$} 
\Text(-7,-24)[cb]{$i_2$}
\Text(2,-31)[cb]{$j_2$}
\Text(50,-9)[cb]{$i_3$}
\Text(50,1)[cb]{$j_3$}
\Text(62,-3)[cb]{$\mu$}
\LongArrow(10,20)(20,10)\Text(24,15)[cb]{$p_1$}
\end{picture}
&=&
\left(
\scalebox{0.75}{
\begin{picture}(60,35)(-10,-2)
\ArrowLine(0,20)(17,3)
\ArrowLine(17,3)(40,3)
\ArrowLine(12,0)(-5,15)
\ArrowLine(-5,-15)(12,0)
\ArrowLine(17,-3)(0,-20)
\ArrowLine(40,-3)(17,-3)
\Text(-2,22)[cb]{$j_1$} 
\Text(-9,15)[cb]{$i_1$} 
\Text(-9,-20)[cb]{$j_2$}
\Text(-2,-27)[cb]{$i_2$}
\Text(48,-9)[cb]{$j_3$}
\Text(48,1)[cb]{$i_3$}
\end{picture}
}
-
\scalebox{0.75}{
\begin{picture}(60,35)(-10,-2)
\ArrowLine(17,3)(0,20)     
\ArrowLine(40,3)(17,3)
\ArrowLine(-5,15)(12,0)
\ArrowLine(12,0)(-5,-15)
\ArrowLine(0,-20)(17,-3)
\ArrowLine(17,-3)(40,-3)
\Text(-2,22)[cb]{$i_1$} 
\Text(-9,15)[cb]{$j_1$} 
\Text(-9,-20)[cb]{$i_2$}
\Text(-2,-27)[cb]{$j_2$}
\Text(48,-9)[cb]{$i_3$}
\Text(48,1)[cb]{$j_3$}
\end{picture}
}
\right)
\frac{\ri\,\gs}{\sqrt{2}}\,p_1^\mu
=
\left(
  \delta^{i_1}_{j_2}\delta^{i_2}_{j_3}\delta^{i_3}_{j_1}
- \delta^{i_1}_{j_3}\delta^{i_2}_{j_1}\delta^{i_3}_{j_2}
\right)
\frac{\ri\,\gs}{\sqrt{2}}\,p_1^\mu\,.
\\[0ex]\nonumber
\label{colourflow rules}
\eeqar
In all Feynman rules the colour part is described by products of
Kronecker $\delta$s, and therefore in the colour-flow representation
the colour structure of the amplitude can be simply obtained as a
linear combination of all possible products of Kronecker $\delta$s
carrying the colour indices of the external particles.  For a process
with $k$ external gluons and $m$ external quark--antiquark pairs the
amplitude takes the simple form:
\begin{equation}
A^{\alpha_1,\dots,\alpha_n}_{\beta_1,\dots,\beta_n} = 
\sum_{P(1,\dots,n)} \delta^{\alpha_1}_{\beta_1}\!\cdots\delta^{\alpha_n}_{\beta_n}\,
A_{1,\dots,n},
\qquad
n = k+m,
\label{colourflow amplitude}
\end{equation}
where {in general} all $n!$ permutations $P(1,\dots,n)$ of the indices
$\beta_1,\ldots,\beta_n$ have to be considered.

In a framework based on colour-dressed amplitudes, the colour indices
of the external particles would be fixed and at each branch, given the
colour of the generating currents, all colour configurations (3 for
quarks or antiquarks, 9 for gluons) for the generated current
would be computed.  Many of them are zero, and the others differ
just by simple factors.  Since in this approach one would define and
compute {unnecessarily} many currents, we decided to follow a
different strategy.

Instead of assigning an explicit colour to the currents, we assign them a 
``colour structure", which is a product of Kronecker $\delta$s.
In order to understand how these structures look like, let us first 
consider the external currents for quarks, antiquarks, and gluons:
\begin{equation}
\begin{picture}(34,10)(0,-2)
\ArrowLine(0,0)(20,0)
\GCirc(20,.5){1.5}{0}
\Text(-5,-5)[cb]{$\beta$}
\Text(27,-3)[cb]{$i$}
\end{picture}
= \,
u_\lambda(p)\,\delta_\beta^i,
\qquad\quad
\begin{picture}(34,10)(0,-2)
\ArrowLine(20,0)(0,0)
\GCirc(20,.5){1.5}{0}
\Text(-5,-3)[cb]{$\alpha$}
\Text(27,-5)[cb]{$j$}
\end{picture}
= \,
\bar{v}_\lambda(p)\,\delta_j^\alpha,
\qquad\quad
\begin{picture}(34,10)(0,-2)
\Gluon(0,0)(20,0){2.5}{3}
\GCirc(20,.5){1.5}{0}
\Text(-5,0)[cb]{\small{$\beta$}}
\Text(-5,-7)[cb]{\small{$\alpha$}}
\Text(27,3)[cb]{$i$}
\Text(26,-9)[cb]{$j$}
\end{picture}
= \;
\epsilon_\lambda(p)\,\delta_\beta^i\,\delta^\alpha_j\,,
\label{coloured currents}
\end{equation}
where $\alpha$ and $\beta$ are the colour indices of the external
particles while $i$ and $j$ are ``open" colour indices, which, during
the recursive construction of internal currents, are contracted with
the indices of the Feynman rules of \refeq{colourflow rules}.  In the
recursive procedure these contractions generate products of $\delta$s:
some of them carry indices of external particles only, as in
\refeq{colourflow amplitude}, and some others involve the open indices
of the generated current.  For example, the combination of an external
quark with colour structure {$\delta_{\beta_1}^{i_1}$} with an
external gluon with colour structure
$\delta_{\beta_2}^{i_2}\,\delta^{\alpha_2}_{j_2}$ produces, according
to the Feynman rules of \refeq{colourflow rules}, a quark with two
possible colour structures:
$\delta_{\beta_1}^{\alpha_2}\delta_{\beta_2}^{i}$ and
$\delta_{\beta_2}^{\alpha_2}\delta_{\beta_1}^{i}$.  In both structures
the first $\delta$ carries just external indices $\alpha_1$, $\beta_1$,
$\beta_2$, while the second one contains also the {open index {$i$} of
  the generated quark current.

The resulting colour structures for the off-shell currents are in
complete correspondence to the colour structure in (\ref{colourflow
  amplitude}) for the full amplitude.  The indices of the
$\delta$-structures of a particular off-shell current are given by the
colour indices $\alpha_i,\beta_j$ of the external particles generating
the current and by potential open indices for the generated particle:
no open indices for a colour-neutral particle, one for a
quark/antiquark, two for a gluon.  Therefore, in general, the colour
structure of a gluon current, obtained from $k$ external gluons and
$n-k$ external quark--antiquark pairs, takes the form
\beq
\begin{picture}(65,40)(-5,-2)
\Gluon(24,0)(49,0){1.5}{6}
\GCirc(50,0){1.5}{0}
\Gluon(15,25)(20,0){1.5}{7}
\Gluon(2,18)(20,0){1.5}{7}
\DashCArc(20,0)(22,110,125){1.5}
\ArrowLine(10,-3)(-5,-8)
\ArrowLine(10, 3)(-5, 8)
\DashCArc(20,0)(22,172.5,187.5){1.5}
\ArrowLine(2,-18)(13,-7)
\ArrowLine(15,-25)(18,-10)
\DashCArc(20,0)(22,235,250){1.5}
\GCirc(20,0){10}{.8}
\Text(13,28)[cb]{\scriptsize $\alpha_{\!1}$} 
\Text(21,28)[cb]{\scriptsize $\beta_1$} 
\Text(-5,16)[cb]{\scriptsize $\alpha_k$} 
\Text( 1,21)[cb]{\scriptsize $\beta_k$} 
\Text(-14,5)[cb]{\scriptsize $\alpha_{k\!+\!1}$} 
\Text(-14,-11)[cb]{\scriptsize $\alpha_{n}$} 
\Text(-2,-26)[cb]{\scriptsize $\beta_{k\!+\!1}$} 
\Text(15,-36)[cb]{\scriptsize $\beta_{n}$} 
\Text(58, 3)[cb]{$i$} 
\Text(58,-9)[cb]{$j$} 
\end{picture}
\quad\to\quad
\delta^{\alpha_{1}}_{\beta_1}\!\cdots\delta^{\alpha_{n-1}}_{\beta_{n-1}}\,
\delta^{\,i}_{\beta_{n}}\,\delta^{\alpha_{n}}_{j}\,,
\eeq

\vspace{1.1cm}
\noindent
where permutations $P(1,\dots,n+1)$ of the indices
$\beta_1,\ldots,\beta_n,j$ on the right-hand side correspond to
different currents.  For the colour structure of a quark (antiquark)
current, obtained from $k$ external gluons, $n-k-1$ external
quark--antiquark pairs and an additional quark (antiquark) we have
\beq
\qquad\quad
\begin{picture}(65,40)(-5,-2)
\ArrowLine(30,0)(50,0)
\GCirc(50,0){1.5}{0}
\Gluon(15,25)(20,0){1.5}{7}
\Gluon(2,18)(20,0){1.5}{7}
\DashCArc(20,0)(22,110,125){1.5}
\ArrowLine(10,-3)(-5,-8)
\ArrowLine(10, 3)(-5, 8)
\DashCArc(20,0)(22,172.5,187.5){1.5}
\ArrowLine(2,-18)(13,-7) 
\ArrowLine(15,-25)(18,-10)
\DashCArc(20,0)(22,235,250){1.5}
\GCirc(20,0){10}{.8}
\Text(13,28)[cb]{\scriptsize $\alpha_{\!1}$} 
\Text(21,28)[cb]{\scriptsize $\beta_1$} 
\Text(-5,16)[cb]{\scriptsize $\alpha_k$} 
\Text( 1,21)[cb]{\scriptsize $\beta_k$} 
\Text(-14,5)[cb]{\scriptsize $\alpha_{k\!+\!1}$} 
\Text(-14,-11)[cb]{\scriptsize $\alpha_{n\!-\!1}$} 
\Text(-2,-26)[cb]{\scriptsize $\beta_{k\!+\!1}$} 
\Text(15,-36)[cb]{\scriptsize $\beta_{n}$} 
\Text(57,-3)[cb]{$i$} 
\end{picture}
\;\,\to\;\,
\delta^{\alpha_{1}}_{\beta_1}\!\cdots\delta^{\alpha_{n-1}}_{\beta_{n-1}}\,
\delta^{\,i}_{\beta_{n}},
\qquad\qquad
\begin{picture}(65,40)(-5,-2)
\ArrowLine(50,0)(30,0)
\GCirc(50,0){1.5}{0}
\Gluon(15,25)(20,0){1.5}{7}
\Gluon(2,18)(20,0){1.5}{7}
\DashCArc(20,0)(22,110,125){1.5}
\ArrowLine(10,-3)(-5,-8)
\ArrowLine(10, 3)(-5, 8)
\DashCArc(20,0)(22,172.5,187.5){1.5}
\ArrowLine(2,-18)(13,-7) 
\ArrowLine(15,-25)(18,-10)
\DashCArc(20,0)(22,235,250){1.5}
\GCirc(20,0){10}{.8}
\Text(13,28)[cb]{\scriptsize $\alpha_{\!1}$} 
\Text(21,28)[cb]{\scriptsize $\beta_1$} 
\Text(-5,16)[cb]{\scriptsize $\alpha_k$} 
\Text( 1,21)[cb]{\scriptsize $\beta_k$} 
\Text(-14,5)[cb]{\scriptsize $\alpha_{k\!+\!1}$} 
\Text(-10,-11)[cb]{\scriptsize $\alpha_{n}$} 
\Text(-2,-26)[cb]{\scriptsize $\beta_{k\!+\!1}$} 
\Text(15,-36)[cb]{\scriptsize $\beta_{n\!-\!1}$} 
\Text(57,-3)[cb]{$j$} 
\end{picture}
\;\,\to\;\,
\delta^{\alpha_{1}}_{\beta_1}\!\cdots\delta^{\alpha_{n-1}}_{\beta_{n-1}}\,
\delta^{\alpha_{n}}_{j},\!\!\!\!\!\!\!\!\!
\eeq

\vspace{1.1cm}
\noindent
where again permutations $P(1,\dots,n)$ {of
  $\beta_1,\ldots,\beta_n$} in the first case and of
$\beta_1,\ldots,\beta_{n-1},j$ in the second correspond to different
currents.

We can easily distinguish two parts in the colour structure: the
``open part", which is always present in coloured currents, contains
one (for quarks and antiquarks) or two (for gluons) $\delta$s with
open colour indices $i$ and/or $j$, while the ``saturated part", which
is absent in external currents, is a product of $\delta$s with only
external colour indices.  Only the open parts of the structures play
an active role in the combination of currents, while the saturated
parts of the generating currents simply multiply to give the saturated
part of the generated current.  In our code these two parts are
represented by integer numbers, based on a binary notation.

Moving from colour-dressed to structure-dressed currents reduces
already the number of currents. For example, there are 9
colour-dressed currents for a gluon generated from the currents of a
quark and an antiquark (although many of them vanish), while we have
just one structure-dressed current.  A further optimisation can be
achieved {introducing a colour label and} giving the same colour label
to currents differing just by a colour factor (due to subsequent
multiplication of different colour coefficients).  In the example
considered above of an external quark combined with an external gluon,
the coefficients of the two possible colour structures
$\delta_{\beta_1}^{\alpha_2}\delta_{\beta_2}^{i}$ and
$\delta_{\beta_2}^{\alpha_2}\delta_{\beta_1}^{i}$ differ only by a
factor $-1/N_c$.  For structure-dressed currents this {label} is
easily introduced, together with the corresponding colour factors,
already in the initialisation phase and can be used in the production
phase of the code to compute just one of the currents with the same
colour label.

\section{Electroweak corrections to $\PZ+2\,$jets production at the LHC}
\label{ppZjj}

As a first example for the application of the code {\recola}, we
consider the EW corrections to the dominant partonic channels
contributing to the process $\Pp\Pp\to\PZ+2\,$jets.

\subsection{Details of the calculation}
\label{ppZjj details}

At leading order (LO) in perturbation theory, the production of a
$\PZ$~boson at the LHC in association with a pair of hard jets is
governed by the partonic subprocesses
\begin{eqnarray}
   \Pq_i\,\Pg&\to&\Pq_i\,\Pg\,\PZ\,,\label{eq:born_qg}\\
   \Pq_i\,\bar{\Pq}_i&\to&\Pq_j\,\bar{\Pq}_j\,\PZ,
   {\qquad q_i,q_j=\Pu,\Pc,\Pd,\Ps,\Pb}
\label{eq:born_qq},
\end{eqnarray}
and their crossing-related counterparts. Since we neglect flavour
mixing as well as finite-mass effects for the light quarks, the LO
amplitudes do not depend on the quark generation, and the
contributions of the various generations to the cross section differ
only by their parton luminosities.  While the mixed quark--gluon
(gluonic) channels \refeq{eq:born_qg} contribute to the cross section
exclusively at order $\mathcal{O}(\alpha\alphas^2)$, the four-quark
channels \refeq{eq:born_qq} develop LO diagrams of strong
as well as of EW nature leading to contributions of order
$\mathcal{O}(\alpha\alphas^2)$, $\mathcal{O}(\alpha^2\alphas)$, and
$\mathcal{O}(\alpha^3)$ to the cross section. Representative Feynman
diagrams are shown in \reffi{fig:LOdiags}. If standard experimental
acceptance cuts are applied (see \refse{se:SMinput} for the
specification of our cuts), the mixed quark--gluon channels clearly
dominate over the four-quark channels, {with the subprocesses
  $\Pu\Pg\to \Pu\Pg\PZ$ and $\Pd\Pg\to \Pd\Pg\PZ$ contributing $\sim
  70\%$ and the complete class \refeq{eq:born_qg} of partonic
  subprocesses contributing $\sim 80\%$ to the total cross section}.%
\footnote{In a scenario where vector-boson fusion kinematics is
  imposed, the dominance of the gluonic channels \refeq{eq:born_qg}
  shrinks. Requiring, for instance,  two
  tagging jets in opposite hemispheres with rapidity difference
  $|y_{\mathrm{j}_1}-y_{\mathrm{j}_2}|>4$, the gluonic channels
  contribute only $\sim 65\%$.}
Therefore as a first step towards a complete NLO calculation of
EW effects in $\PZ+2$\,jets we here calculate EW
corrections to the gluonic channels \refeq{eq:born_qg}.

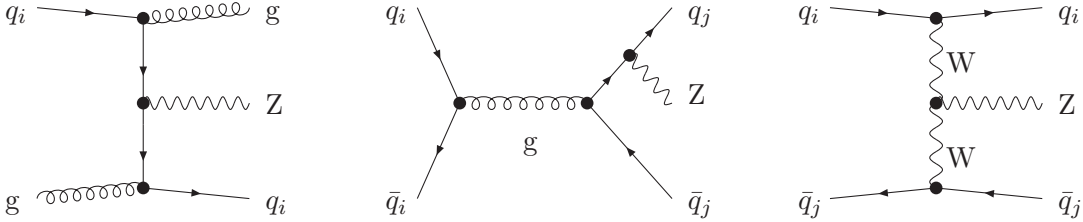
\begin{figure}
\centerline{
{\unitlength .8pt 
\begin{picture}(125,100)(0,0)
\SetScale{.8}
\ArrowLine( 15,95)( 65,90)
\Gluon( 15, 5)( 65,10){3}{7}
\ArrowLine( 65,10)(115, 5)
\Gluon(115,95)( 65,90){3}{7}
\ArrowLine( 65,90)( 65,40)
\ArrowLine( 65,40)( 65,10)
\Photon(115,50)( 65,50){3}{7}
\Vertex(65,90){3}
\Vertex(65,10){3}
\Vertex(65,50){3}
\put(  0,90){$q_i$}
\put(  0,0){$\Pg$}
\put(123,90){{$\Pg$}}
\put(123,0){{$q_i$}}
\put(123,45){{$\PZ$}}
\SetScale{1}
\end{picture}
}
\hspace*{3em}
{\unitlength .8pt 
\begin{picture}(145,100)(0,0)
\SetScale{.8}
\ArrowLine( 15,95)( 35,50)
\ArrowLine( 35,50)( 15, 5)
\ArrowLine(95,50)(115,72.5)
\ArrowLine(115,72.5)(135,95)
\ArrowLine(135, 5)(95,50)
\Gluon( 35,50)(95,50){3}{7}
\Photon(115,72.5)(135,50){3}{4}
\Vertex( 35,50){3}
\Vertex(95,50){3}
\Vertex( 115,72.5){3}
\put(  0,90){$q_i$}
\put(  0,0){$\bar{q}_i$}
\put(143,90){{$q_j$}}
\put(143,0){{$\bar{q}_j$}}
\put(143,50){{$\PZ$}}
\put( 65,28){{$\Pg$}}
\SetScale{1}
\end{picture}
\hspace*{3em}
{\unitlength .8pt 
\begin{picture}(125,100)(0,0)
\SetScale{.8}
\ArrowLine( 15,95)( 65,90)
\ArrowLine( 65,10)( 15, 5)
\ArrowLine(115, 5)( 65,10)
\ArrowLine( 65,90)(115,95)
\Photon( 65,10)( 65,90){3}{7}
\Photon(115,50)( 65,50){3}{7}
\Vertex(65,90){3}
\Vertex(65,10){3}
\Vertex(65,50){3}
\put(  0,90){$q_i$}
\put(  0,0){$\bar{q}_j$}
\put(123,90){{$q_i$}}
\put(123,0){{$\bar{q}_j$}}
\put(123,45){{$\PZ$}}
\put( 70,65){{$\PW$}}
\put( 70,20){{$\PW$}}
\SetScale{1}
\end{picture}
}
}
}
\caption{From left to right: Sample tree diagrams for the QCD
  contributions to $\Pq_i\,\Pg\to\Pq_i\,\Pg\,\PZ$ and to
  $\Pq_i\,\bar{\Pq}_i\to\Pq_j\,\bar{\Pq}_j\,\PZ$, and the EW
  contributions to $\Pq_i\,\bar{\Pq}_j\to\Pq_i\,\bar{\Pq}_j\,\PZ$.} 
\label{fig:LOdiags}
\end{figure}

\subsubsection{General setup}

In our calculation we describe potentially resonant $\PZ$-boson
propagators appearing in loop diagrams (see \reffi{fig:NLOdiags} left
for a sample diagram) by attributing a complex mass
\begin{equation}
\mu_{\PZ}^2 = \MZ^2 -\ri \MZ \GZ
\end{equation}
to internal $\PZ$ bosons.  To this end, we consistently use the
complex-mass scheme \cite{Denner:1999gp,Denner:2005fg,Denner:2006ic}
where $\mu^2_\PW$ and $\mu^2_\PZ$ are defined as the poles of the
$\PW$- and $\PZ$-boson propagators in the complex plane. On the other
hand, the external $\PZ$ boson is treated as a stable final-state
particle with its invariant mass being fixed to $\MZ$, where $\MZ^2$
is the real part of the complex pole of the $\PZ$-boson propagator.
The pole values $M_V$ and $\Gamma_V$ ($V=\PW,\PZ$) for the mass and
width of the $\PW$ and $\PZ$~boson are related to the on-shell results
$M_V^{\mathrm{OS}}$ and $\Gamma_V^{\mathrm{OS}}$ obtained from the LEP
and Tevatron experiments by \cite{Bardin:1988xt}
\beq\label{eq:m_ga_pole}
M_V = M_V^{\OS}/ \sqrt{1+(\Ga_V^{\OS}/M_V^{\OS})^2}\,, \qquad \Ga_V =
\Ga_V^{\OS}/ \sqrt{1+(\Ga_V^{\OS}/M_V^{\OS})^2}\,.
\eeq

For the definition of the electromagnetic coupling constant $\alpha$
we adopt the $G_\mu$ scheme, {i.e.\ }we fix the value of
$\alpha$ via its tree-level relation with the Fermi constant $G_\mu$:
\beq\label{eq:Gmu}
 \alpha_{\GF} = \frac{\sqrt{2}\GF\MW^2}{\pi}\left(1-\frac{\MW^2}{\MZ^2}\right).
\eeq
Compared to the Thomson-limit definition of $\alpha$, the definition
of $\alpha_{\GF}$ in the $G_\mu$-scheme incorporates effects of the
renormalisation-group running from the scale $Q^2=0$ to the scale
$Q^2=M_\PW^2$. In addition NLO corrections involving logarithms of
light quark masses are avoided as such contributions do not enter the
muon decay.

\subsubsection{Virtual corrections}
The virtual corrections involve $\mathcal{O}(300)$ diagrams per
partonic channel, including 20 pentagon and 71 box
contributions.%
\footnote{While \recola\ does not use Feynman diagrams, we give these
  numbers as a measure of the complexity of the process.}  The most
complicated topologies are given by pentagons involving 5-point
functions of rank up to $r=4$ (see \reffi{fig:NLOdiags} right for a
sample diagram). {The virtual} amplitude is calculated using the
't~Hooft--Feynman gauge. The calculation of the tensor integrals is
performed employing recursive numerical reduction to scalar integrals
based on
\citeres{Denner:2002ii,Denner:2005nn,Beenakker:1990jr,Denner:1991qq,Denner:2010tr}.
Numerical instabilities from small Gram determinants are avoided by
resorting to various expansion algorithms for the problematic momentum
configurations \cite{Denner:2005nn}. Both, in the case of UV
divergences as well as in the case of infrared (IR) divergences,
dimensional regularisation is applied to extract the corresponding
singularities. The EW sector of the SM is renormalised using an
on-shell prescription for the $\PW$- and $\PZ$-boson masses in the
framework of the complex-mass scheme \cite{Denner:2005fg}.  As the
coupling $\alpha_{G_\mu}$ is derived from $\MW$, $\MZ$ and $G_\mu$,
its counterterm inherits a correction term $\Delta r$ from the weak
corrections to muon decay\cite{Dittmaier:2001ay}.
\begin{figure}
\centerline{
{\unitlength .8pt 
\begin{picture}(125,100)(0,0)
\SetScale{.8}
\Gluon( 15,95)( 55,75){3}{5}
\Gluon( 15, 5)( 55,25){3}{5}
\ArrowLine(55,75)(55,25)
\ArrowLine(55,25)(95,25)
\ArrowLine(95,25)(95,75)
\ArrowLine(95,75)(55,75)
\Photon(95,25)(135,25){3}{4}
\ArrowLine(135,25)(165,5)
\ArrowLine(165,45)(135,25)
\Photon(95,75)( 135,95){3}{4}
\Vertex(55,25){3}
\Vertex(55,75){3}
\Vertex(95,25){3}
\Vertex(95,75){3}
\Vertex(135,25){3}
\put(  0,90){$\Pg$}
\put(  0,0){$\Pg$}
\put(143,90){{$\PZ$}}
\put(173,0){{$q_i$}}
\put(173,40){{$\bar{q}_i$}}
\put(113,5){{$\PZ$}}
\SetScale{1}
\end{picture}
}
\hspace*{9em}
{\unitlength .8pt 
\begin{picture}(125,100)(0,0)
\SetScale{.8}
\ArrowLine( 15,95)( 45,90)
\Gluon( 45,10)( 15, 5){3}{3}
\ArrowLine( 105,90)( 135,95)
\Gluon( 135,5)( 105, 10){3}{3}
\ArrowLine(45,90)(45,10)
\ArrowLine(45,10)(105,10)
\ArrowLine(105,10)(105,50)
\ArrowLine(105,50)(105,90)
\Photon(45,90)(105,90){3}{7}
\Photon(135,50)( 105,50){3}{4}
\Vertex(45,90){3}
\Vertex(45,10){3}
\Vertex(105,90){3}
\Vertex(105,10){3}
\Vertex(105,50){3}
\put(  0,90){$q_i$}
\put(  0,0){$\Pg$}
\put(143,90){{$q_i$}}
\put(143,0){{$\Pg$}}
\put(143,45){{$\PZ$}}
\put(65,100){{$\PZ/\gamma$}}
\SetScale{1}
\end{picture}
}
}

\caption{Left: Box diagram involving a potentially resonant
  $\PZ$-boson propagator.  Right: Pentagon diagram involving a 5-point
  tensor integral of rank 4.}
\label{fig:NLOdiags}
\end{figure}
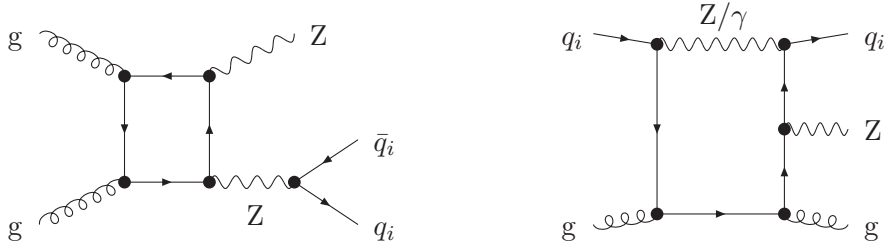

For virtual NLO contributions the finite top-quark mass affects
partonic channels involving external bottom quarks in a different way
than channels with external quarks of the first two generations. While
the top-quark mass is properly taken into account in closed fermion
loops, finite top-quark-mass effects constrained to diagrams with
external bottom quarks are neglected  (${\raisebox{.7em}{\tiny $(-)$}\hspace{-.83em}\Pb}\Pg\to{\raisebox{.7em}{\tiny $(-)$}\hspace{-.83em}\Pb}\Pg\PZ$ and
$\Pb\Pbbar\to\Pg\Pg\PZ$ are suppressed by the bottom PDFs,
$\Pg\Pg\to\Pb\Pbbar\PZ$ contributes about 1\% at LO).

\subsubsection{Real corrections}
\label{real_corr}

The EW real corrections to the subprocess \refeq{eq:born_qg}
are induced by photon Bremsstrahlung and given by
\beq
   \Pq_i\,\Pg\to\Pq_i\,\Pg\,\PZ\,\gamma\,.\label{eq:real_qg}
\eeq
Emission of a soft or a collinear photon from an external quark leads
to IR divergences which are regularised dimensionally. If an IR-save
event definition is used, the final-state singularities cancel with
corresponding IR poles from the virtual corrections. For the
initial-state singularities this cancellation is incomplete but the
remnant can be absorbed into a redefinition of the quark distribution
function. Technically we make use of the Catani--Seymour dipole
formalism as formulated in \citere{Catani:1996vz}, which we
transferred in a straightforward way to the case of dimensionally
regularised photon emission.

In addition to the singularities from soft and collinear photon
emission we face a further source of IR divergences originating from a
soft final-state gluon (see \citere{Denner:2009gj}).  Isolated soft
gluons do not pose any problem as they do not pass our selection cuts
because the requirement of two hard jets is not fulfilled.  However,
in IR-safe observables quarks, and thus all QCD partons, have to be
recombined with photons if they are sufficiently collinear.  Thus, if
a soft gluon is collinear to a photon, it still passes the selection
cuts if recombined with the collinear photon, giving rise to a
soft-gluon divergence {that would be cancelled by the virtual QCD
  corrections to $\PZ+1\,\mathrm{jet}+\gamma$ production.}  Following
\citeres{Denner:2009gj,Denner:2010ia} we eliminate this singularity by
discarding events which contain a jet consisting of a hard photon
recombined with a soft parton $a$ ($a=q_i,\bar{q}_i,\Pg$) taking the
photon--jet energy fraction $z_{\gamma}=
E_{\gamma}/(E_{\gamma}+E_{a})$ as a discriminator.  Photonic jets with
$z_\gamma$ above a critical value $z_\gamma^{\mathrm{cut}}$ are
attributed to the process $\Pp\Pp\to \PZ+1\,\textrm{jet}+\gamma$ and
therefore they are excluded.  However, this event definition is still
not IR-save because the application of the $z_\gamma$-cut to
recombined quark--photon jets spoils the cancellation of final-state
collinear singularities with the IR-divergences from the virtual
corrections. This is cured by absorbing the left-over singularities
into the measured quark--photon fragmentation function
\cite{Glover:1993xc,Buskulic:1995au}.

\subsubsection{Implementation}
In order to ensure correctness of our results, and in particular {of}
the calculation with \recola, we have performed two independent
calculations which we find to be in mutual agreement.  While the first
one applies the technique of recursive amplitude generation as
described in \refse{bulba}, the second one relies on the conventional
Feynman-diagrammatic approach.

In the first {calculation} the amplitude generator \recola\ provides
the 
Feynman amplitudes.  For the
evaluation of the tensor integrals \recola\ is interfaced with the
{\sc Fortran} library {\sc Collier} \cite{collier}. To this end {\sc
  Collier} has been extended by an efficient algorithm for building up
the tensor integrals from the recursively calculated Lorentz-invariant
coefficient functions. The phase-space integration is performed by
means of a generic in-house Monte-Carlo generator \cite{Motz}
following the multi-channel sampling approach.

The second code uses {\sc FeynArts}~3.2 \cite{Hahn:2000kx,Hahn:2001rv}
and {\sc FormCalc}~3.1 \cite{Hahn:1998yk} for the generation and
simplification of the Feynman amplitudes. For the numerical evaluation
the amplitudes are translated into the Weyl--van~der~Waerden formalism
\cite{Dittmaier:1998nn} using the program {\sc Pole}
\cite{Accomando:2005ra}. The tensor integrals are again evaluated by
{\sc Collier} which by itself provides two independent implementations
of all its building blocks. Finally, the phase-space integration is
performed with the multi-channel generator {\sc Lusifer}
\cite{Dittmaier:2002ap}.

\subsubsection{Accuracy and efficiency of \recola}

In this section we estimate the accuracy and efficiency of the purely
numerical algorithm \recola\ by comparing with the code {\sc Pole} which
is based on algebraically generated analytical expressions.

\begin{table}
\begin{center}
\renewcommand{\arraystretch}{1.2}
\renewcommand{\tabcolsep}{4pt}
   \begin{tabular}{|c|c|c|c|c|c|c|c|}
     \hline 
     Process class &  virtual $[{\rm fb}]$ & $|{\mathrm{R\over P}}-1|$$[\%]$ & real $[{\rm fb}]$& $|\mathrm{R\over P}-1|$$[\%]$ &  $\delta\sigma^{\NLO}_{\EW}~[{\rm fb}]$ & $|\mathrm{R\over P}-1|$$[\%]$  \\\hline 
$\Pq\Pg\to \Pq\Pg \PZ$,              & {$-14463 \pm 10 $} &\multirow{2}{*}{$0.3 \pm 0.2$} & {$-825 \pm 9$} &\multirow{2}{*}{$2 \pm 3 $}  &{$-15288 \pm 13$}&\multirow{2}{*}{$0.3 \pm 0.2 $} \\
$\bar{\Pq} \Pg\to \bar{\Pq} \Pg \PZ$ & $-14499 \pm 27$    && $-841 \pm 22$ && $ -15340 \pm 35$ & \\ 
\hline 
\multirow{2}{*} {$\Pq\bar{\Pq} \to \Pg\Pg\PZ$} & {$-1395 \pm 2 $} &\multirow{2}{*}{$0.8 \pm 0.5$} & {$\phantom{-}118 \pm 1$}&\multirow{2}{*}{$0.01 \pm 1 $} & {$-1277 \pm 2$}&\multirow{2}{*}{$0.9 \pm 0.6$} \\
                                                                 & $-1406 \pm 7 $  && $\phantom{-}118 \pm 1$   && $-1288 \pm 7$ & \\
\hline  
\multirow{2}{*}{$\Pg\Pg \to \Pq\bar{\Pq}\PZ$} &{$ -1024 \pm 2$} &\multirow{2}{*}{$0.5 \pm 0.4$} & {$-186 \pm 1$}&\multirow{2}{*}{$0.7 \pm 0.9 $}  & {$-1210 \pm 2$}&\multirow{2}{*}{$0.3 \pm 0.3$}  \\
                                                         & $-1018\pm 3$    && $ -187 \pm 1$  && $-1206 \pm 3 $ &\\
\hline

   \end{tabular}
\end{center}
  \caption{Comparison of numerical results from {\recola} (upper
    numbers) and {\sc Pole} (lower numbers) for the NLO contribution
    $\delta\si^{\NLO}_{\EW}$ to the total cross section of various
    partonic process classed (summed over $q=\Pu,\Pd$). In addition
    to the complete NLO correction we separately
    give the finite virtual and real corrections and the relative
    differences between {\recola} and  {\sc Pole}.}
  \label{tab:compare-pole-recola}
\end{table}

In \refta{tab:compare-pole-recola} we compare   results obtained with
{\recola} (upper numbers) 
and {\sc Pole} (lower numbers) for the NLO contribution
$\delta\sigma^{\NLO}_{\EW}$ to the total cross section for various
classes of 
partonic channels. We further display
separate results for the finite virtual corrections including the
integrated dipoles (virtual) and for the finite real corrections
including the dipole-subtraction terms (real). In addition we also
present the relative deviation $|\mathrm{R/P}-1|$ between the results of 
\recola\ ($\mathrm{R}$) and {\sc Pole}\ ($\mathrm{P}$). For the
results obtained by {\recola} we have requested $5\times 
10^6$ accepted events in the case of the virtual corrections and
$10^8$ accepted events in the case of the real corrections, while
(roughly by a factor of 10) lower statistics  
has been used for the calculation with {\sc Pole}. 
The results in \refta{tab:compare-pole-recola} demonstrate that our
two independent calculations agree with each other within the Monte
Carlo errors in the per-mille range.

A more detailed comparison can be obtained by comparing the weights at
individual phase space points. For
$10^6$ Monte-Carlo generated phase-space points we have compared the
pure virtual contribution
$2\Re(\mathcal{M}^*_{\mathrm{LO}}\delta\mathcal{M}_{\mathrm{NLO}})$
to the squared matrix element (with divergences omitted) calculated by
{\recola} and {\sc Pole} for the partonic process $\Pu\Pg\to\Pu\Pg\PZ$.  
In \reffi{fig:comparison} we show the integrated fraction of the $10^6$
phase-space points for which the agreement $|\mathrm{R/P}-1|$ between {\recola}
and {\sc Pole} is worse than  $\Delta$. We find a typical agreement of
$10^{-11}-10^{-14}$, 
with less than $1\%$ of the phase-space points showing an agreement
worse than $10^{-7}$ and less than $0.02\%$ showing an agreement
worse than $10^{-5}$.   
\begin{figure}
  \begin{center}
  \scalebox{0.8}{\input{./plot_acc.tex}}
  \end{center}
  \caption{Level of agreement of
    $2\Re(\mathcal{M}^*_{\mathrm{LO}}\delta\mathcal{M}_{\mathrm{NLO}})$ 
  between {\recola} and {\sc Pole} for  $\Pu\Pg\to\Pu\Pg\PZ$.  
  The plot shows the probability for an agreement worse than $\Delta$  
  for $10^6$ phase-space points generated by the Monte Carlo.}     
\label{fig:comparison}
\end{figure}
The result of this test of the precision of the code is similar to the
one performed by the {\sc OpenLoops} collaboration
\cite{Cascioli:2011va} which also uses the tensor-integral library
{\sc Collier}. 
Note, however, that we compare two independent
codes for the calculation of the tensor coefficients which are based on two
entirely different algorithms. Furthermore the distribution of the phase-space
points is in our case determined from the
multi-channel Monte Carlo generator adapted to the peaking structure
of the underlying process. 

Finally, we give some details on timing and the amount of memory required.
The evaluation of the spin- and colour summed one-loop matrix elements
takes about 30 ms per 
phase-space point for  $\Pu\Pg\to\Pu\Pg\PZ$ (or any other partonic
process considered in this paper) on a single Intel
i7-2720QM core with {\tt gfortran 
4.6.1}. The size of the executable on disk is about 3\,MB. The matrix
elements are constructed during the run and the required memory
depends strongly on the size of the dynamically generated internal
arrays and thus on the 
considered process. For $\Pp\Pp\to\PZ+2\,$jets memory is no issue.
More complicated processes will be considered in the future.

\subsection{Numerical results}
\label{ppZjj results}

\subsubsection{Input parameters and selection cuts}
\label{se:SMinput}

We use the following set of input parameters \cite{Beringer:1900zz},
\begin{equation}
       \begin{array}[b]{rclrcl}
          \GF & = & 1.1663787 \times 10^{-5} \GeV^{-2}, \quad\\
          \MW^{\OS} & = & 80.385\GeV,\quad &
          \Gamma_\PW^{\OS} & = & 2.085\GeV, \\
          \MZ^{\OS} & = & 91.1876\GeV, &
          \Gamma_\PZ^{\OS} & = & 2.4952\GeV, \\
          M_\PH & = & 125\GeV,\quad &
          m_\Pt & = & 173.2\GeV
       \end{array}
       \label{eq:SMinput}
\end{equation}
with the value for the top-quark mass taken from
\citere{Lancaster:2011wr}. The pole masses $M_{\PW,\PZ}$ and widths
$\Gamma_{\PW,\PZ}$ entering our calculation are obtained from the
stated on-shell values $M^{\mathrm{OS}}_{\PW,\PZ}$ according to
\refeq{eq:m_ga_pole}. The electromagnetic coupling constant
$\alpha_{G_\mu}$ is determined from $\GF$ via \refeq{eq:Gmu}. The
CKM~matrix only appears in loop amplitudes and is set to {unity.}
       
For the prediction of the hadronic $\Pp\Pp\to \PZ+2\,${jets} cross
section the partonic cross sections have to be convoluted with the
corresponding parton distribution functions (PDFs).  Since our
calculation does not take into account NLO QCD effects, we
consistently resort to LO PDFs, using the LHAPDF implementation of the
central MSTW2008LO PDF set \cite{Martin:2009iq}. From there we infer
the value
\beq\label{eq:als}
\alphas^\LO(\MZ)=0.1394
\eeq
for the strong coupling constant. We identify the QCD factorisation
scale $\mu_{\mathrm{F}}$ and the renormalisation scale
$\mu_{\mathrm{R}}$ choosing
\beq
  \mu_{\mathrm{F}}=\mu_{\mathrm{R}}=\MZ\,.
\eeq
Note {that} the choice of the scales $\mu_{\mathrm{F},\mathrm{R}}$ as
well as the actual value for the strong coupling $\alphas$ plays a
minor role for our numerical analysis of EW radiative corrections in
{\refse{se:distributions}}. We focus on the relative importance of the
NLO EW corrections considering the ratio
$\sigma_\EW^\NLO/\sigma^\LO$ from which the $\alphas$- and the scale
dependence drop out.

For the jet-reconstruction we use the anti-$k_\mathrm{T}$ clustering
algorithm \cite{Cacciari:2008gp} with separation parameter $R=0.4$.
For our scenario with {exactly two partons and one potential photon}
in the final state this simply amounts to recombining the photon with
a parton $a$ {if} $R_{a\ga} =
\sqrt{(y_{a}-y_{\ga})^2+\phi_{a\ga}^2}<R$. Here $y = \frac{1}{2} \ln
[(E + p_\RL)/(E - p_\RL)]$ is the particle's {rapidity} with $E$
denoting its energy and $p_{\RL}$ its three-momentum component along
the beam axis, and $\phi_{a\ga}$ is the azimuthal angle between the
the photon and the parton $a$ in the plane transverse to the beam
axis. In case of recombination, the resulting photon--parton jet is
subjected to the cut $z_\gamma=E_\gamma/(E_\gamma+E_a)< 0.7$ in order
to distinguish between $\PZ+2$ jets and {$\PZ+1\,\mathrm{jet}+\ga$}
production as explained in \refse{real_corr}. After a possible
recombination, we require two hard jets with
\beq
   \label{eq:cuts}
   p_{\mathrm{T},\mathrm{jet}}>25\GeV, \qquad |y_\mathrm{jet}|<4.5
\eeq
{for the final event.}

\subsubsection{Results}
\label{se:distributions}
In this section we present results for the total cross section and
various differential distributions using the numerical input
parameters and acceptance cuts introduced above.  The total cross
section and its composition at LO for the {$8\TeV$ LHC} is shown in
\refta{tab:x-section} where the absolute and relative contributions of
the partonic channels are listed.
\begin{table}
\begin{center}
\renewcommand{\arraystretch}{1.2}
   \begin{tabular}{|c|c|c|c|c|}
     \hline
&&&&\\[-.4cm]
Process class  & $\sigma^\LO~ [\rm{ pb}] $  & $ 
\sigma^\LO/\sigma^\LO_{\rm tot}$ $[\%]$   & $\sigma_\EW^\NLO~ [\rm{ pb}] $  & $ 
\frac{\sigma_\EW^\NLO}{\sigma^\LO}-1$ $[\%]$ \\[+.2cm]
     \hline $\Pu\Pg\to \Pu\Pg \PZ$, $\Pd\Pg\to \Pd\Pg \PZ$,& 
\multirow{2}{*}{1324.1(2)} & \multirow{2}{*}{68.79} 
&\multirow{2}{*}{1308.8(2)}  & \multirow{2}{*}{$-1.16$}\\
      $\Pubar \Pg\to \Pubar \Pg \PZ$, $\Pdbar \Pg\to \Pdbar \Pg \PZ$ &&&&\\
     \hline $\Pu\Pubar \to \Pg\Pg Z$, $\Pd\Pdbar \to \Pg\Pg Z$ & 128.84(2) & 6.69 
&{127.56(2)}  &$- 0.99$\\
     \hline $\Pg\Pg \to \Pu\Pubar Z$, $ \Pg\Pg \to \Pd\Pdbar Z$ &  87.40(2) & 4.54 
& 86.18(2) & $-1.37$ \\
     \hline $\Pu\Pu'\to \Pu\Pu' \PZ$, $\Pd\Pd'\to \Pd\Pd' \PZ$, & 
\multirow{2}{*}{88.41(2)} & \multirow{2}{*}{4.59} &\multirow{2}{*}{---}  
& \multirow{2}{*}{---}\\
      $\Pubar\Pubar'\to \Pubar\Pubar' \PZ$, $\Pdbar\Pdbar'\to \Pdbar\Pdbar' \PZ$ &&&&\\
     \hline $\Pu\Pubar\to \Pu'\Pubar' \PZ$,  $\Pd\Pdbar\to \Pd'\Pdbar' \PZ$, & 
\multirow{2}{*}{87.98(2)} & \multirow{2}{*}{4.57} &\multirow{2}{*}{---}  
& \multirow{2}{*}{---}\\
     $\Pu\Pubar'\to \Pu\Pubar' \PZ$,  $\Pd\Pdbar'\to \Pd\Pdbar' \PZ$ &&&&\\
     \hline $\Pu\Pubar\to \Pd\Pdbar \PZ$, $\Pd\Pdbar\to \Pu\Pubar \PZ$, & 
\multirow{2}{*}{16.566(3)} & \multirow{2}{*}{0.86} 
&\multirow{2}{*}{---}  & \multirow{2}{*}{---}\\
     $\Pu\Pubar'\to \Pd\Pdbar' \PZ$,  $\Pd\Pdbar'\to \Pu\Pubar' \PZ$ &&&&\\
     \hline $\Pu\Pd\to \Pu'\Pd' \PZ$, $\Pubar\Pdbar\to \Pubar'\Pdbar' \PZ$,& 
\multirow{2}{*}{111.74(3)} & \multirow{2}{*}{5.81} 
&\multirow{2}{*}{---}  & \multirow{2}{*}{---}\\
      $\Pu\Pd\to \Pu\Pd \PZ$, $\Pubar\Pdbar\to \Pubar\Pdbar \PZ$&&&& \\
     \hline $\Pu\Pdbar\to \Pu'\Pdbar' \PZ$, $\Pubar \Pd\to \Pubar'\Pd' \PZ$, & 
\multirow{2}{*}{79.70(2)} & \multirow{2}{*}{4.14} &\multirow{2}{*}{---}  
& \multirow{2}{*}{---}\\
     $\Pu\Pdbar\to \Pu\Pdbar \PZ$, $\Pubar \Pd\to \Pubar \Pd \PZ$&&&&\\
     \hline
     \hline gluonic &  1540.4(2)  &  80.02 & 1522.5(2) & $-1.16$ \\
     \hline four-quark & 384.41(4)   & 19.98 &---&--- \\
     \hline sum & 1924.8(2)  & 100.00 &---&---\\
     \hline
   \end{tabular}
\end{center}
  \caption{Composition of the LO cross section for 
    $\Pp\Pp \to \PZ+2\,$jets at the $8\TeV$ LHC. In the first column the
    partonic processes are listed, where $u,u'$ denote the up-type
    quarks $\Pu,\Pc$ and   
   $d,d'$ the down-type quarks $\Pd,\Ps,\Pb$. The second column
   provides the corresponding cross section where the numbers in
   parenthesis give the integration error on the last digit. The third
   column contains the relative contribution to the total cross section
   in percent. In the fourth column we list the NLO EW cross section for
   the gluonic channels and in the last column the relative EW corrections.}
  \label{tab:x-section}
\end{table}
In the lower part of \refta{tab:x-section} we provide the contribution
to the total cross section of partonic processes with external gluons
(gluonic) and {of the} four-quark processes (four-quark). We find the
total cross section dominated by processes with external gluons, in
particular by the quark-gluon induced processes (69\%).  We also
provide the NLO cross section and the relative EW corrections for the
gluonic channels in the last two columns of \refta{tab:x-section}. For
our set of cuts, they range between ${-1.0\%}$ and ${-1.4\%}$ for the
different gluonic channels.

In the following we present results for distributions at LO and NLO
for the gluonic channels only. Although these channels dominate the
total cross section we emphasise that there are certain phase-space
regions where the relative importance of the four-quark processes is
enhanced and these channels and the corresponding EW corrections
should not be neglected to describe {$\Pp\Pp\to \PZ+2$ jets} properly.

For each distribution we provide two plots: the upper panels
show the LO and NLO prediction for the differential cross section
while the lower panels show the NLO result normalised to the LO
result.
 \begin{figure}
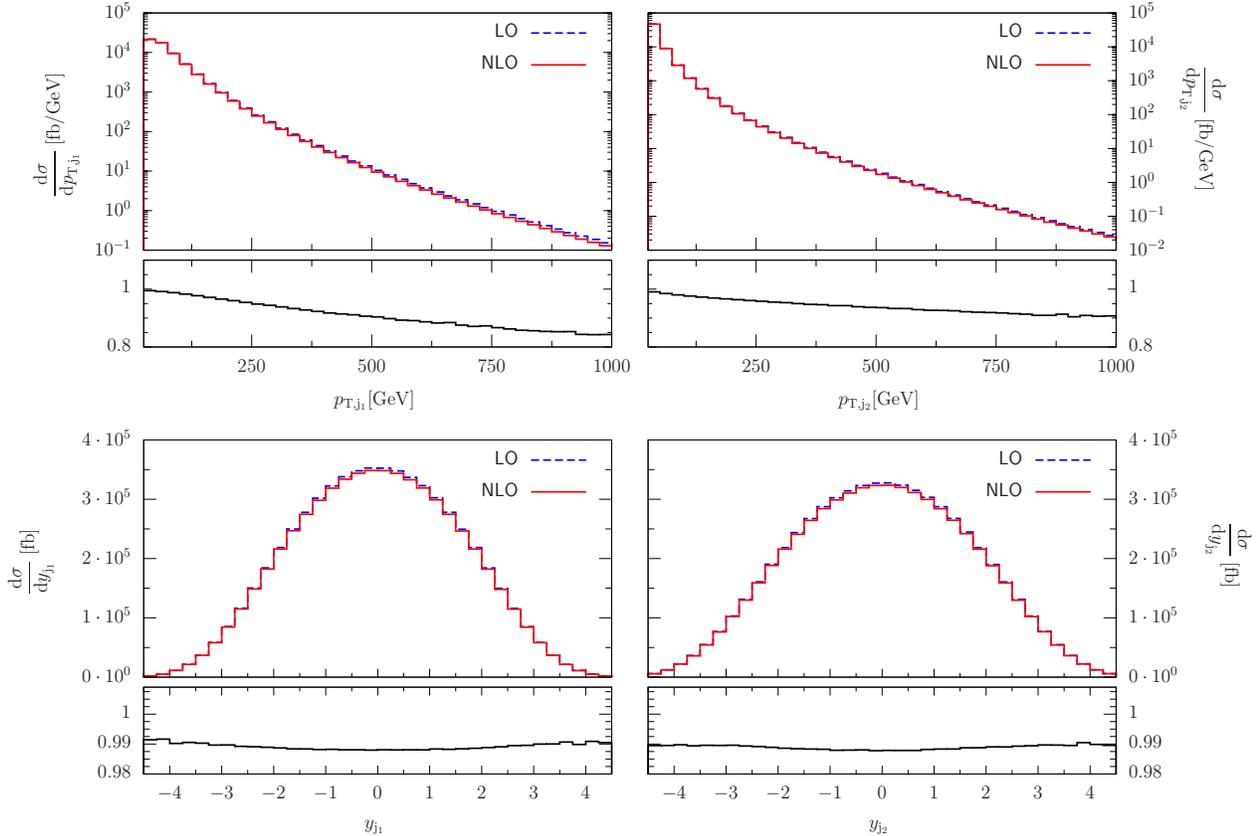

 \begin{center}
  \scalebox{0.5}{\input{LHC8_ZJJ_pTjetH.tex}} 
\hspace*{1mm}
  \scalebox{0.5}{\input{LHC8_ZJJ_pTjetS.tex}}\\[12mm]
  \scalebox{0.5}{\input{LHC8_ZJJ_etajetH.tex}} 
\hspace*{1mm}
  \scalebox{0.5}{\input{LHC8_ZJJ_etajetS.tex}}
\end{center}
\caption{Distributions of the transverse momentum and the rapidity 
  of the harder jet ${\rm j_1}$ and the softer jet ${\rm j_2}$ at the 
  $8\TeV$ LHC at LO (blue, dashed) and NLO (red, solid). 
  The lower panels show the ratio of the NLO distribution
  over the LO distribution.}
\label{fig:num1}
\end{figure}
In \reffi{fig:num1} we {present} results for the differential cross
section as a function of the transverse momentum and the {rapidity}
for the harder { jet $\rm {j_1}$ and softer jet $\rm {j_2}$,}
respectively.  Both transverse momentum distributions show steep
slopes over six orders of magnitude in the {displayed} $\pT$ range.
The EW corrections lower the LO prediction, and their relative size
grows in absolute value with increasing transverse momentum due to the
well-known EW Sudakov logarithms
\cite{Ciafaloni:1998xg,Kuhn:1999de,Fadin:1999bq,Denner:2000jv}.  For
transverse momenta of the softer jet $p_{\rm T,j_2} {\simeq 1\TeV}$
the impact of EW corrections amounts up to $-10\%$ relative to the
Born approximation, while for {transverse momenta of the harder jet}
$p_{\rm T,j_1}\simeq 1\TeV$ the EW effects are of the order of
$-15\%$. {Since, the {rapidity} distributions are not sensitive to
  Sudakov logarithms, the corresponding EW corrections are flat and
  around $-1\%$ as for the total cross section.}

\begin{figure}[t]
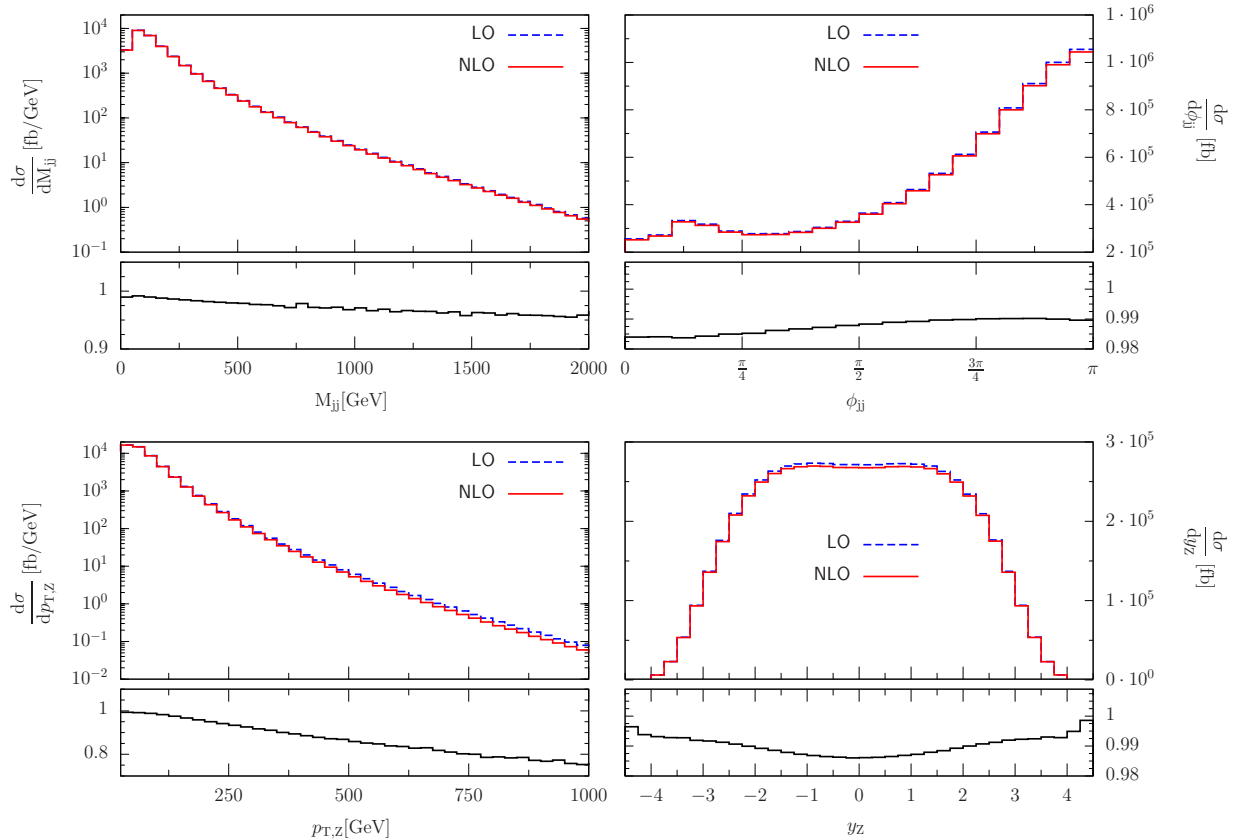

 \begin{center}
  \scalebox{0.5}{\input{LHC8_ZJJ_Mjj.tex}} 
\hspace*{1mm}
  \scalebox{0.5}{\input{LHC8_ZJJ_Azimuth.tex}}\\[12mm]
  \scalebox{0.5}{\input{LHC8_ZJJ_pTZ.tex}} 
\hspace*{1mm}
  \scalebox{0.5}{\input{LHC8_ZJJ_etaZ.tex}}
\end{center}
\caption{Distributions of the di-jet invariant mass, the relative 
  azimuthal angle between the two jets, the transverse momentum and
  the rapidity of the Z boson at the $8\TeV$ LHC at LO (blue, dashed)
  and NLO (red, solid).  The lower panels show the ratio of the NLO
  distribution over the LO distribution.}
\label{fig:num2}
\end{figure}
The differential cross section as a function of the di-jet invariant
mass and {as a function of} the transverse momentum of the Z boson is
shown on the left-hand side in \reffi{fig:num2}.  For both
distributions we find the expected dependence on the Sudakov
logarithms, although the sensitivity in the di-jet {invariant-mass
  distribution} is less pronounced than in the $p_{\rm
  T,Z}$-distribution.  For $M_{\rm jj} \simeq 500\GeV$ the EW
corrections are of the order of $-2\%${; they} amount up to $-4 \%$
for $M_{\rm jj} {\simeq} 2\TeV$. The transverse momentum distribution
of the Z boson receives large corrections, {from} $-15 \%$ for $p_{\rm
  T,Z}\simeq500\GeV$ to $-25\%$ for {$p_{\rm T,Z}\simeq 1\TeV$.}  On
the upper right-hand side in \reffi{fig:num2} we present the
differential distribution of the relative azimuthal angle $\phi_{\rm
  jj}$ between the two jets.  {The $\phi_{\rm jj}$-distribution shows
  that the two jets are preferably back-to-back in the transverse
  plane} and {that} the EW corrections lower the differential cross
section by $1{-}1.5\%$. They induce a shape change at the permille
level relative to the {LO} approximation.  The {rapidity} distribution
of the Z boson is {depicted} in the lower right-hand side of
\reffi{fig:num2}.  In the central region $|y_\PZ|<2$, where most
of the Z bosons are produced, the EW corrections lower the LO cross
section by $1{-}1.5\%$ while for $|y_\PZ|>2$ their effect drops to
the permille level.

\section{Conclusions}
\label{conclusions}
The full exploitation of the Large Hadron Collider relies on precise
theoretical predictions. To this end QCD and electroweak
next-to-leading order corrections have to be calculated for many
processes involving many particles in the final state. This requires 
efficient and reliable automatic tools. 

In this paper we have presented \recola, a {\sc Fortran90} code
for the REcursive Computation of One-Loop Amplitudes. It uses methods
based on Dyson--Schwinger equations to calculate the coefficients of
all tensor integrals appearing in a one-loop amplitude recursively.
The tensor integrals can then be evaluated with efficient numerically
stable techniques. The algorithm has been implemented for the full
electroweak Standard Model, including counterterms and rational terms,
but could be generalised to more complicated theories in a
{straightforward} way.  The implementation supports the complex-mass
scheme and is thus applicable to processes involving intermediate
unstable particles.  For the treatment of colour we {have developed} a
new {recursive} algorithm based on colour structures that
naturally appear in the colour-flow representation.

As a first application of \recola, we have calculated the electroweak
corrections to the dominant partonic channels in $\PZ+2\,$jets
production at the LHC. The results have been verified with an
independent calculation based on Feynman-diagrammatic methods. For a
typical set of cuts, the electroweak corrections are negative at the
level of {one} percent, but become sizeable where large energy scales
are relevant. However, in general and in particular for large energies
of the jets, a meaningful prediction requires the inclusion of
next-to-leading-order corrections to all partonic channels. This
will be pursued in a forthcoming publication.

\subsection*{Acknowledgements}
We are grateful to S. Pozzorini for many useful discussions and to
A. M\"uck for help concerning {\sc Pole}.
This work was supported in part by the Swiss National Science
Foundation (SNF) under contract 200021-126364, by the Deutsche
Forschungsgemeinschaft (DFG) under reference number DE~623/2-1,
and by the Helmholtz alliance ``Physics at the Terascale".

\end{document}